# Active Expressions

## Basic Building Blocks for Reactive Programming

Stefan Ramson[a] and Robert Hirschfeld[a]

a    Hasso Plattner Institute, University of Potsdam, Germany

**Abstract**    Modern software development without reactive programming is hard to imagine. Reactive programming favors a wide class of contemporary software systems that respond to user input, network messages, and other events.

While reactive programming is an active field of research, the implementation of reactive concepts remains challenging. In particular, change detection represents a hard but inevitable necessity when implementing reactive concepts. Typically, change detection mechanisms are not intended for reuse but are tightly coupled to the particular change resolution mechanism. As a result, developers often have to re-implement similar abstractions. A reusable primitive for change detection is still missing.

To find a suitable primitive, we identify commonalities in existing reactive concepts. We discover a class of reactive concepts, *state-based reactive concepts*. All state-based reactive concepts share a common change detection mechanism: they detect changes in the evaluation result of an expression.

On the basis of the identified common change detection mechanism, we propose *active expressions* as a reusable primitive. By abstracting the tedious implementation details of change detection, active expressions can ease the implementation of reactive programming concepts.

We evaluate the design of active expressions by re-implementing a number of existing state-based reactive concepts using them. The resulting implementations highlight the expressiveness of active expressions.

Active expressions enable the separation of essential from non-essential parts when reasoning about reactive programming concepts. By using active expressions as a primitive for change detection, developers of reactive language constructs and runtime support can now focus on the design of how application programmers should be able to react to change. Ultimately, we would like active expressions to encourage experiments with novel reactive programming concepts and with that to yield a wider variety of them to explore.

# The Art, Science, and Engineering of Programming

Active Expressions

# 1 Introduction

It is hard to imagine modern software development without reactive programming. Many contemporary software systems are inherently reactive: a Web application needs to react to messages from the server, editing a cell in a spreadsheet might change other cells that depend on the edited cell's value, and embedded software reacts to signals of the hardware. Even though reactive programming is widely-used, plain imperative programming is still the norm due to the intrinsic nature of the underlying machine model. However, empirical studies hint at potential advantages of reactive programming over imperative approaches. Studies of non-programmer's solutions reveal that reactive programming reflects natural reasoning about problems [27]. Further studies examine program comprehension in graphical animations and interactive applications. The studies reveal that reactive programming improves the correctness of comprehending reactive behavior compared to traditional object-oriented programming (OOP) solutions [28]. For those reasons, reactive programming concepts often seem a desirable abstraction over imperative environments.

Although there are various incarnations of the reactive programming paradigm, their implementations usually consist of two parts: *detection of change* and *reaction to change*. The reaction part defines how to propagate detected changes through the system. The reaction is usually seen as the heart of a reactive concept. Most design decisions regard this part, as it is the part of actual, visible functionality. The detection part is responsible for detecting changes and events. In contrast to reaction, the detection remains hidden to the application programmer. Also, the detection part serves no inherent, concept-specific purpose, but instead is conceptually exchangeable. Therefore, this part is often perceived as an implementation detail. However, the detection has a major impact on practical implementations. This issue becomes especially apparent when dealing with a change detection mechanism beyond manual event-emitting, for example detecting when a constraint expression becomes unsatisfied in a scenario that respects object-oriented encapsulation [8]. Ideally, even complex change detection remains invisible and does not introduce additional friction into systems or workflows. More often, the detection part becomes a limiting factor when the used implementation does not cover all important cases. Covering the missing cases might range from being tedious and time-consuming to impossible.

Implementing complex detection behavior on top of an imperative execution environment typically involves either adapting the virtual machine (VM), using meta programming and reflection, alternating the compilation process of a program, or imposing conventions on the users of the concept. Each of these options comes with its own limitations. For example, one drawback of a VM-based approach is that, although based on a particular host-language, the extension will only work on the customized VM. However, in many languages that are of interest to both industry and academia, applications typically have to work on a variety of client VMs.

A major problem is that one can hardly reuse existing implementations of reactive concepts. This is due to the fact that existing implementations are typically highly focused on a very specific use case. As a consequence, the detection mechanism and the reactive behavior are tightly coupled together in order to maximize performance and expressiveness. As the existing solutions are not intended for reuse, developers often have to start from scratch again.

Change detection is a tedious but inevitable necessity when implementing reactive programming concepts. Developers often have to unnecessarily re-create similar abstractions. This issue draws the attention of application programmers and language implementors away from the interesting reactive parts to the limitations of their practical implementation. Well-chosen detection primitives could provide a common ground for reactive programming concepts to build upon and thereby help to accelerate the development of novel reactive programming concepts.

Thus, we propose *active expressions* as a common foundation for reactive programming concepts that is explicitly designed to relieve developers of the recurring chore of change detection. Developers may specify the state they want to monitor for changes using expressions available in the underlying host language. Whenever the evaluation result of an expression changes, for example due to an assignment to a variable referenced by the expression, active expressions recognize this as a change. Once a change is detected, dependent components are notified and may invoke desired behavior in reaction to that change. An implementation of active expressions is available in JavaScript. To demonstrate how to make use of active expressions as a reactive primitive, we generalize the implementation of four different reactive programming concepts, each relying on state change detection.

**Contributions**

In this paper, we make the following contributions towards easier development of novel reactive programming concepts:

- We identify *state-based reactive concepts* as a subset of reactive concepts that share a common change detection mechanism.
- We propose the design of *active expressions*, a primitive reactive concept that acts as unified foundation for change detection of state-based reactive concepts.
- We provide a prototypical implementation of active expressions in JavaScript incorporating multiple implementation strategies for change detection. We discuss conceptual limitations and runtime overhead of the implementation strategies.
- We exemplify how to implement existing reactive concepts using active expressions.

**Outline**

The remaining of this paper is structured as follows. In section 2, we provide an overview of several state-based reactive concepts and identify their common underlying structure. In section 3, we present the design of active expressions. We first introduce the guiding principles and goals of our design. After that, we explain how to detect state changes in an object-oriented environment in a way that respects encapsulation. Then, section 4 describes three implementation strategies for active expressions and discusses their conceptual limitations and runtime overhead. In section 5, we exemplify how to use active expressions to re-implement existing reactive programming

**Active Expressions**

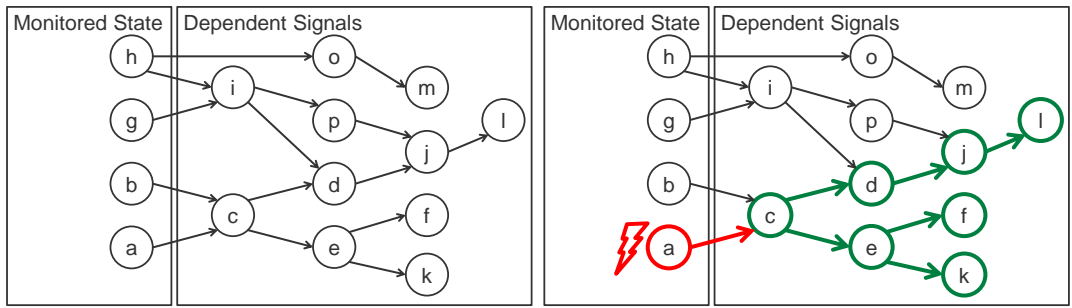

(a) Signal implementations maintain a dependency network of signals and basic variables.

(b) Changing an ordinary variable causes all signals in its convex hull to update.

**Figure 1** State monitoring and change propagation for signals

concepts. Furthermore, we compare our active expression-based implementations against reference implementation. Next, section 6 relates active expressions to existing approaches. Finally, section 7 presents our conclusions and discusses future work.

## 2 A Recurring Reactive Pattern

In the past few years, reactive programming has become of particular interest for researchers in the language design community due to its relevance for contemporary software. Thus, many different kinds of reactive concepts have been proposed or are still under active research. In the following, we highlight specific reactive programming concepts. For each example, we briefly explain its working principle and how it detects change. Finally, we identify a commonality in the structure of the presented concepts and describe the class of *state-based reactive concepts*.

**Signals** Signals are time-varying values [6]. They can be declared like any other variable by providing an expression:

```
1  var a = 5, b = 6;
2  signal c = a + b; // c = 11
3
4  a = 10; // c = 16
```

However, instead of performing the assignment once, the declaration of a signal introduces a functional dependency, as shown in line 2. As a result, the value of c is re-computed according to its production rule by the underlying reactive framework whenever a or b change, as in line 4. Thus, the relation c = a + b is always true.

To keep track of the functional dependencies in a program the reactive framework maintains an acyclic dependency graph as exemplified in figure 1a. Additionally, all ordinary variables referenced by a signal are continuously monitored. Whenever the reactive framework detects an assignment to a monitored variable, the change of this variable propagates through the dependency network, ultimately updating all dependent signals, as shown in figure 1b.

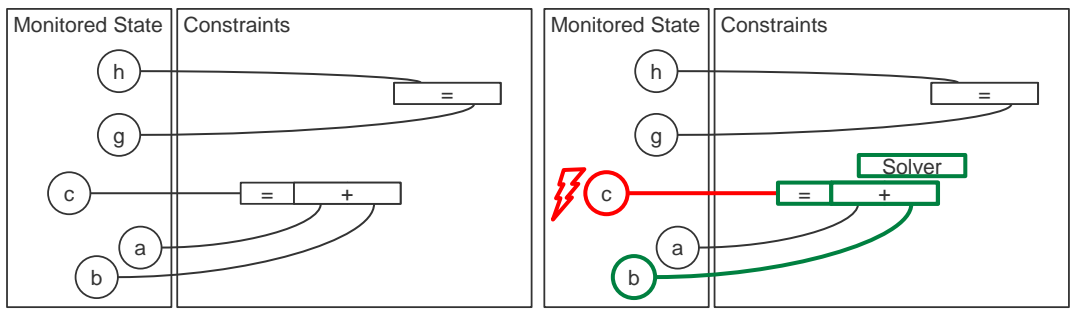

**(a)** Constraints maintain desired relations between variables.

**(b)** Assigning new values to variables might invalidate constraints. In response, constraint solvers adjust variables to keep the desired relations satisfied.

■ **Figure 2** State monitoring and change propagation for constraints

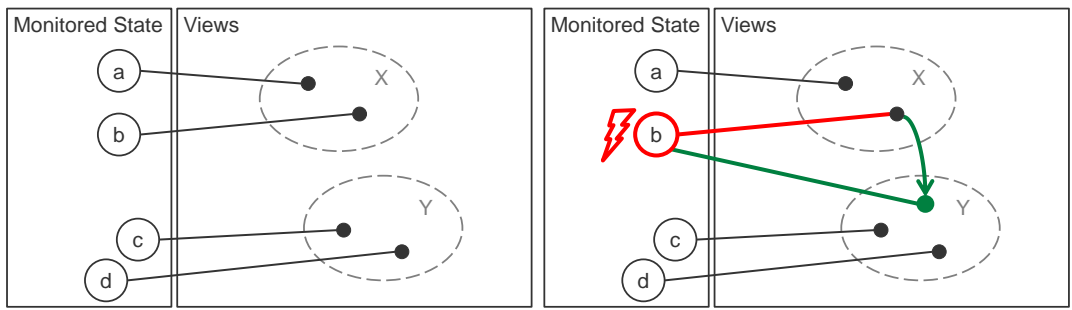

**(a)** Views contain objects depending on their properties.

**(b)** Views react to changes to object properties and update their content in accordance to these changes.

■ **Figure 3** State monitoring and change propagation for reactive object queries

**Constraints** Constraint-imperative programming [10, 14] and object constraint programming [9] aim to integrate constraints into imperative languages. Constraints are relations between objects that should hold. When specifying a constraint, specialized constraint solvers are used to adapt variables in a way to fulfill the given condition:

```
1  var a = 1, b = 1, c = 1;
2  always: a + b == c; // a = 1, b = 1, c = 2
3  c = 5; // a = 1, b = 4, c = 5
```

During further execution, imperative code might again invalidate the given condition, as exemplified in line 3. Thus, the variables referenced in the constraint are monitored for changes by the underlying reactive framework, as shown in figure 2a. When a change invalidates a constraint, the system uses constraint solvers again to maintain a consistent system state according to the specified constraints, as seen in figure 2b.

**Reactive Object Queries** Reactive object queries [19] apply reactivity to data structures, namely sets. Instead of manually constructing and maintaining a set of objects,

**Active Expressions**

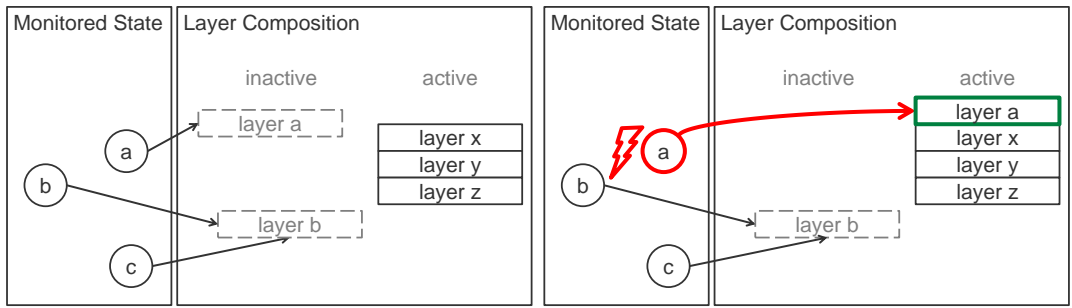

**(a)** The layer composition stack determines in which order active class extensions apply to a method. Implicitly activated layers are active while a given condition holds.

**(b)** Changing the program state might affect the layer composition

■ **Figure 4** State monitoring and change propagation for implicit layer activation one can use a query to automatically construct a set of instances of a certain class that fulfill a given condition:

```
1  var taskA = new Task('Write paper'),
2      taskB = new Task('Provide code example'),
3      todos = select(Task, t => !t.done()); // todos = [taskA, taskB]
4  taskB.finish(); // todos = [taskA]
```

The resulting set acts similar to a view in conventional databases: it automatically updates whenever the program state changes. As a result, the set always maintains consistency with the underlying system state, as exemplified in line 4. To do so, the underlying reactive framework monitors all class instances for changes that could affect the given condition, as shown in figure 3a. When relevant values change, the condition is re-evaluated and the set is updated accordingly, as illustrated in figure 3b.

**Implicit Layer Activation** In context-oriented programming (COP) [15], one can define multiple class extensions in a single unit of modularity, called a *layer*. During the execution of a program this layer can be activated to dynamically apply the class extensions, thus, adapting the behavior of a program. There are multiple means to activate a layer [16]. One of those activation means, implicit layer activation (ILA) [21], has fairly reactive semantics. Using ILA, a layer is not activated or deactivated at a fixed time, but instead is active while a given condition holds:

```
1  var shouldTrace = false;
2  new Layer().refineObject(Networking, {
3    fetch(url) {
4      console.log('fetch ' + url);
5      return proceed(url);
6    }
7  }).activeWhile(() => shouldTrace);
8
9  Networking.fetch('example.com'); // prints nothing
10 shouldTrace = true;
11 Networking.fetch('example.com'); // prints 'fetch example.com'
```

Despite these reactive semantics, ILA is typically implemented in an imperative manner. An underlying system keeps track of all layered methods, such as the method `fetch` adapted in line 3. Then, when calling a layered method, the current layer composition stack is determined [16]. At this very point in time, the context-oriented programming (COP) framework checks the conditions of all implicitly activated layers [1, 25]. If the condition evaluates to true, the method adaptation is taken into account for this method call, as the modified behavior in line 11 illustrates.

Considering the reactive semantics of ILA, a reactive implementation does not seem a stretch [18]: an underlying reactive framework may monitor variables referenced by the given condition, as depicted in figure 4a. When such a variable changes, the condition is re-evaluated and the corresponding layer is activated or deactivated accordingly, as shown in figure 4b.

⁂

All reactive concepts share the same overall structure of *change detection* and *reaction to change*. The latter part, reaction, works quite different for each of the presented concepts as each concept tackles a very specific use case. In contrast to the diversity in their reactions, all presented concepts share a common semantics for change detection: the reactive framework for each of the concepts monitors certain parts of the program state for changes and, whenever a change is detected, reacts to this change in a concept-specific manner. Thus, the presented concepts provide means for developers to specify the state to monitor in form of an *expression*:

| Signals | Constraints | Object Queries | Implicit Layers |
|---|---|---|---|
| **signal** s = `expr`; | **always**: `expr`; | **select**(Class, `expr`); | layer.activeWhile(`expr`); |

Semantically, an underlying reactive framework continuously evaluates the expression. Then, whenever the evaluation result differs from the previous one, a change is detected. For example, an assignment to a local variable might change the evaluation result of a constraint expression from true to false, which is detected by the reactive framework as a change. In reaction to that, the reactive framework uses a constraint solver to satisfy the desired relation. This recurring detection mechanism is not limited to the presented concepts, but also applies to other reactive concepts such as two-way data bindings and incremental lists [22]. Because all these concepts react to changes in the state of a program, we coin this identified class of reactive concepts *state-based reactive concepts*. To be specific, the class of state-based reactive concepts includes all reactive concepts in which *dependencies are specified implicitly as expressions over program state*. To exemplify this definition, the aforementioned constraint example belongs to this class as the underlying system identified the local variable as a dependency from the constraint expression provided by the user. As a counter example, the definition excludes Reactive Extensions [23], because the programmer explicitly constructs the dependency graph. The discovery of this class encourages to think about a state-based reactive concepts in terms of their commonalities rather than their differences. By reifying the common change detection mechanism into a reactive primitive, we can ease the development of novel reactive programming approaches.

## 3 Designing the State-based Reactive Primitive

The overarching design goal is to relieve developers of the recurring chore of state change monitoring. Therefore, our concept, *active expressions,* has to meet the following requirements:

1. Active expressions should ease the detection of state changes for developers by hiding technology-specific implementation details.
2. Concepts built using active expressions should still be able to perform a variety of reactive behavior. Therefore, active expressions should impose as few restrictions and assumptions on the reactive behavior of state-based reactive concepts as possible.
3. A large portion of developers is familiar with object-oriented (oo) languages and a large number of useful code is written in oo languages. Thus, active expressions should integrate well with existing oop languages.

To simplify further decisions, we design active expressions as a state-based reactive concept. By breaking down active expressions into a detection and a reaction part, we can reason about each part independently. The detection part deals with how to specify and detect changes, while the reaction part deals with the propagation of change to other concepts. In the following, we describe the design of each part separately.

### 3.1 Expressions as Abstraction over State

For specifying which part of the system state to monitor for a particular behavior, we have basically two options: let the concept programmer explicitly specify which variables and members to track for change detection, or provide a higher-level abstraction over state. The first option provides fine-grained control over the state to be monitored for change. For example, a trackMember function could install listeners for the change of a given member attribute. To detect changes to the width property of a rectangle object, we would write:

```
trackMember(rectangle, 'width');
```

However, explicit specification provides little improvement over classical mechanisms like reflection. Imagine an alternative implementation of the rectangle object that instead of having a direct width attribute, has an extent attribute of type point. The x attribute of this point is the attribute to detect. Now, to be able to detect all possible changes to rectangle.extent.x, we actually have to track two members:

```
1 trackMember(rectangle, 'extent');
2 trackMember(rectangle.extent, 'x');
```

Additionally, the programmer has now the duty to update the second listener whenever a new point is assigned to the extent attribute of the rectangle. Therefore, this option introduces a lot of manual overhead for programmer even in simple cases. Thus, this option is contrary to requirement 1. Even worse, let assume that the rectangle hides its implementation inside a getter function: rectangle.width(). In this scenario, the programmer needs to know the specific implementation upfront and is forced to circumvent OO encapsulation to listen for changes, thus, additionally contradicting with requirement 3.

A suitable abstraction needs to limit the manual overhead for programmers to fulfill requirement 1. Therefore, we use *expressions* as an abstraction over state. Programmers can provide an expression to describe which part of the system state to monitor. Active expressions will detect a change whenever the *evaluation result of the given expression changes*. To be concrete, we detect an updated result whenever the object identity of the result changes. Consider the previous example of monitoring the width of a rectangle for change:

```
aexpr(() => rectangle.width);
```

The function **aexpr** takes an expression as parameter. Depending on whether the host language supports first-class blocks, programmers provide the expression as a block or use a function as in the example above. In this example, the evaluation result of the expression changes when either the width attribute of the rectangle changes, or when the reference of the variable rectangle changes to a rectangle object of different size. Both cases lead to the detection of a state change.

The introduction of expressions as abstraction over state has a number of advantages: First, the abstraction relieves the programmer of the responsibility of maintaining listeners. Instead, the reactive framework is in charge of maintaining appropriate listeners to detect all relevant state changes. Continuing the previous example, we use the following expression to track the width of an rectangle with an extent attribute:

```
aexpr(() => rectangle.extent.x);
```

In the example, the reactive framework automatically updates its listeners whenever the reference of the variable rectangle or any of the accessed properties changes. Second, in contrast to explicit listeners, expressions only describe the desired part of the system state to monitor, not how to achieve this. Rather than providing a strict step-by-step instruction, the declarative description of the problem is handed to a reactive framework which then determines how to monitor the desired state. As a result, expressions provide flexibility in the monitoring approach. Third, expressions enable the composition of state changes at the level of the host language. In particular, one can use any control structure available in the host language. For example, we can use a combination of a **for**-loop and conditionals to detect changes to the sum of all positive entries in an array:

**Active Expressions**

```
1  aexpr(() => {
2    let sum = 0;
3    for(let value of arr) {
4      if(value >= 0) sum += value;
5    }
6    return sum;
7  });
```

Fourth, expressions enable the reuse of oo abstractions that exist in the host language. Thus, active expressions respect oo encapsulation and polymorphism. Being able to detect changes in the results of functions, we can reformulate the previous example:

```
1  aexpr(() => {
2    return arr
3      .filter(value => value >= 0)
4      .reduce((acc, value) => acc + value, 0);
5  });
```

Respecting oo encapsulation also contributes towards requirement 3. Fifth, expressions are a familiar concept to users of imperative host languages. By using the same notational elements in imperative and change detection code, active expressions reduce the learning overhead for imperative developers. In fact, active expressions only add a single new concept to host languages.

The expressions that define state to be monitored have a number of restrictions:

- The constraint expression should be free of side effects, or, if there are side effects, those should not influence the result of the expression if evaluated multiple times. As an example, one may use benign side effects for caching purposes.
- The result of evaluating the expression should be deterministic. For example, an expression whose value depended on a randomly generated number would not qualify.

### 3.2 Minimal Predefined Reactive Behavior

Requirement 2 specifies that concepts built with active expressions should still be able to perform a variety of reactive behavior. Thus, active expressions have to avoid unnecessary restrictions and assumptions on the behavior of state-based reactive concepts. To fulfill this requirement, we treat the reaction to changes as a point of variation. Users can subscribe a callback to an active expression using its `onChange` method:

```
aexpr(expr).onChange(callback);
```

Whenever the reactive framework detects a change of the evaluation result of the expression, it invokes every subscribed callback. When calling a callback, we pass the new evaluation result of the expression as an argument to the callback:

```
1  var x = 2;
2  aexpr(() => x).onChange(value => console.log('The value of x changed to ', value));
3  x = 5; // prints: The value of x changed to 5
4  x = 42; // prints: The value of x changed to 42
```

By keeping the reactive part of active expressions minimal, programmers may access the change detection abstraction without unnecessary layers of indirection. As a result, state-based reactive concepts built with active expressions can focus on their reaction, while active expressions deal with change detection.

## 4 Implementation

In industry, JavaScript has become the de-facto standard for Web programming with a rapidly growing amount of code that exists in the language.[1] This fact, along with JavaScript's unique design and its execution environment in a Web browser, makes it of great interest to the research community. Many researchers were motivated to revise and adapt useful features of other languages to the domain of Web programming [26, 32]. We implement our prototype of active expressions in JavaScript, because most modern web applications are reactive and, thus, benefit from active expressions.

Resembling the overall structure of reactive programming concepts, our implementation consists of two parts:

1. *Change Detection*: Monitoring the state of the program for changes
2. *Propagation of Change:* Notifying dependent modules about changes

**Propagation of Change**  The second part is straight-forward to implement. Once a change in the evaluation result of an active expression is detected, we first compare the current evaluation result with the previous one, and, if they differ, invoke all callbacks associated with that active expression with the new result as a parameter.

**Change Detection**  For the first part, we have to continuously monitor the given expression for changes. Therefore, our implementation needs to provide a mechanism to inject custom hooks into program execution to signalize state changes. As described in section 1, there are multiple options to introduce the change detection mechanism into a program:

- A customized VM
- Conventions and guidelines for users of active expressions
- Language features such as meta programming and reflection
- A customized compilation process

JavaScript runs in a variety of client VMs. As the VM-based approach requires a customized VM, it is not suited for a practical implementation of active expressions in JavaScript. In contrast, each of the remaining three options enables programmers to run active expressions in any modern JavaScript environment and to use them in a variety of practical applications. Thus, we provide multiple options to detect state changes.

---

[1] http://stackoverflow.com/research/developer-survey-2016 accessed on February 27th 2017

**Active Expressions**

We describe an implementation for each of the remaining three options in the following subsections. Each active expression uses exactly one of these monitoring strategies. Each implementation strategy is available in its respective repository.[2] Afterwards, we compare the properties of the implementation strategies and discuss their limitations in a qualitative analysis. Furthermore, we provide a micro benchmark to analyze the runtime penalties introduced by each strategy. Finally, we draw conclusions from the analysis to provide guidance on the usage of the different implementation strategies.

**4.1 Explicit Notification by Convention**

The first implementation strategy imposes a convention on the usage of active expressions. In particular, the *convention* strategy requires the programmer to explicitly specify at which points during the execution the implementation should check for possible changes. The programmer marks such a point in the program execution by calling the exposed check function:

```
1  var x = 2;
2  aexpr(() => x).onChange(value => console.log('The value of x changed to ', value));
3  x = 5;
4  check(); // prints: The value of x changed to 5
```

At those user-defined points, the system notifies all active expressions that there have been potential changes, and that now is a good time to check whether there are new results. To do so, the system maintains a set of currently enabled active expressions. Any active expression created is automatically added to that set. Optionally, one may provide an iterable over active expressions as a parameter to check. In this case, we only notify the given active expressions about a potential change in their results.

This monitoring strategy does not guarantee to capture all state changes. Instead, the strategy requires careful manual usage or additional code injection mechanisms to not miss important state changes as exemplified in the following:

```
1  var x = 2;
2  aexpr(() => x).onChange(value => console.log('The value of x changed to ', value));
3  x = 5; // undetected state change
4  x = 17;
5  check(); // prints: The value of x changed to 17
```

**4.2 Interpretation and Reflection**

The second monitoring strategy, the *interpretation* strategy, relies on built-in language features for meta programming and reflection. Unfortunately, JavaScript's meta programming facilities are quite limited. However, *property accessors* can intercept get and set operations on object properties with custom code. Thus, we can use property accessors to track a large portion of changes in the program state.

This state monitoring strategy acts in two stages:

---

[2] https://github.com/active-expressions accessed on November 30th 2016

1. When creating a new active expression, we install property accessors at appropriate places.
2. Once a wrapped property is assigned, we notify all depending active expressions.

**Installation of Property Accessors**   To install appropriate property accessors, we first have to identify all objects and members that contribute to the evaluation result of an expression. To do so, we perform a dynamic interpretation of the expression when a new active expression is created. To perform the interpretation, we use Blockly's [3] JavaScript-in-JavaScript interpreter.[4] We customized the interpreter using means of context-oriented programming (COP) [20] to intercept each visit of a property access. We wrap each property accessed during interpretation in a transparent property accessor. If the property is already wrapped, we add the currently interpreted active expression to a set of active expressions associated with this property instead.

For some expressions, it is not sufficient to determine dependencies only once, for example, if the expression contains if-statements or nested object structures. As a consequence, during later execution, assignments to wrapped properties necessitate re-interpretation to update dependencies correctly.

**Explicit Local Scope**   One limitation of the used interpreter is that it relies on explicit access to the local scope of the expression to interpret. JavaScript does not provide computational access to the local scope by default. To solve this issue, we expand all undeclared occurrences of the identifier locals to an object with all locally accessible references. The following JavaScript source code exemplifies this process:

Before transformation

```
1  var alice, bob;
2  aexpr(() => /*...*/, locals);
3  {
4    let carol;
5    aexpr(() => /*...*/, locals);
6  }
```

After transformation

```
var alice, bob;
aexpr(() => /*...*/, { alice, bob });
{
  let carol;
  aexpr(() => /*...*/, { alice, bob, carol });
}
```

To perform this transformation, we make use of the vast build tool environments for JavaScript that emerged over the last couple of years. By the time of writing, many modern JavaScript projects utilize an additional compilation step to ensure compatibility to legacy execution environments while being able to use newest standards and even experimental language features. Thus, we provide our source code transformation as a plugin for babel,[5] the currently most widely-used JavaScript-to-JavaScript compiler. While the babel compiler traverses a program in form of an abstract syntax tree (AST), each plugin may modify or replace AST nodes. As shown in appendix A, about 12.5% of all JavaScript projects on Github created in 2016 are preprocessed with babel as part of their build process. As a consequence, many developers may option-

---

[3] https://developers.google.com/blockly/ accessed on September 28th 2016
[4] https://github.com/NeilFraser/JS-Interpreter accessed on September 28th 2016
[5] https://babeljs.io/ accessed on September 29th 2016

**Active Expressions**

ally include this plugin to simplify the process of providing the necessary references. Alternatively, one may specify the required scope explicitly.

**Handling Write Accesses**   Once we identified relevant dependencies and installed property accessors accordingly during interpretation, normal execution resumes. Using the installed property accessors, we intercept every write access to an object property that is relevant for an active expression. Whenever a new value is assigned to a wrapped property, we notify all active expressions associated with the property that their evaluation result might have changed.

### 4.3 Alternating Compilation

The third implementation strategy, the *compilation* strategy, captures changes to the system state by injecting hooks into the program using a source code transformation. In contrast to the previous strategy which uses the limited built-in language features to detect changes to the system state, this monitoring strategy uses babel to alternate the compilation process.

This state monitoring strategy acts in three stages:

1. We inject hooks into the source code at compile time.
2. When creating a new active expression, we identify the parts of the state it depends on.
3. Once we detect a change in the system state, we notify all depending active expressions.

**Injection of Monitoring Code**   Only a limited number of language concepts can actually access program state in a way relevant to active expressions. Among these state-modifying concepts are assignments and read accesses to object members and variables. In order to detect changes to the program state, we make these concepts computationally accessible. To do so, whenever we visit a state-modifying or -accessing node during AST traversal, we replace that node with an appropriate function call. In addition to providing the needed hooks into program execution, these functions are designed to keep the original semantics intact, thus, making the rewritten program unaware of the source transformation. The transformation automatically imports the required functions in a non-conflicting manner.

To exemplify the process of the source code transformation, consider the following instance method `equals` of the class `Vector`:

```
1  class Vector2 {
2    // ...
3    equals(vector) {
4      return this.x == vector.x &&
5        this.y == vector.y;
6    }
7  }
```

In this example, we identify four occurrences of object member accesses: **this**.x, vector.x, **this**.y, and vector.y. As shown in figure 5a, these occurrences are represented as Mem-

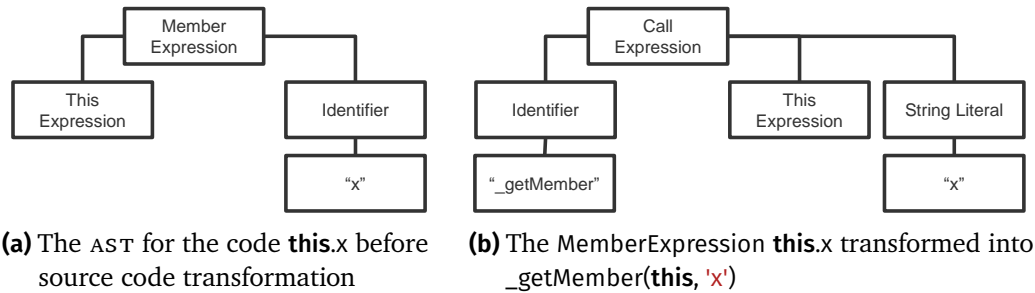

**(a)** The AST for the code **this**.x before source code transformation

**(b)** The MemberExpression **this**.x transformed into _getMember(**this**, 'x')

■ **Figure 5** Two ASTs representing the same code fragment before and after our source code transformation.

berExpression nodes in the corresponding AST. The source code transformation needs to rewrite every MemberExpression node it visits. To do so, we replace the occurrence of a MemberExpression with a CallExpression, as exemplified in figure 5b. In addition, the called function is added as an additional import with a unique identifier. The following code shows the resulting source code after repeating the same procedure for each of the four member accesses:

```
1  import {getMember as _getMember} from 'active—expression—rewriting';
2
3  class Vector2 {
4    // …
5    equals(vector) {
6      return _getMember(this, 'x') == _getMember(vector, 'x') &&
7        _getMember(this, 'y') == _getMember(vector, 'y');
8    }
9  }
```

Analogous to the exemplified transformation of MemberExpression nodes, we perform similar transformations for AssignmentExpressions and CallExpressions on members. Furthermore, we identify and rewrite accesses to global variables and local variables with their respective scope.

We apply the presented transformation statically at compile time. One downside of this approach is that we cannot determine upfront, which portion of the source code will be relevant for active expressions. As a result, the programmer has to decide which modules should be transformed and which should not. Transformed code interacts seamlessly with non-transformed code, because all wrappers are completely transparent. However, non-transformed parts of the source code will not trigger active expressions.

**Expression Analysis**    With the hooks set up, we can create the correct dependencies between program state and active expressions at runtime. Whenever a new active expression is created, we analyze which parts of the program state affect its evaluation result. To do so, we simply execute the expression with a special flag indicating its analysis. During this analysis, we intercept all read accesses. In addition to performing the read operation, we associate the current active expression with the read part of

**Active Expressions**

the system state, either a member with its object or a variable with its scope. Thus, after the analysis of the expression, the active expression is associated with all parts of the program state that might affect the evaluation result of the expression. As with the previous strategy, assignments to dependencies necessitate re-analyzing the expression to correctly handle for example conditionals and nested object structures.

**Handling Write Accesses**   With the dependencies identified, during ordinary code execution, we intercept every write access to a variable or an object member. In addition to the write behavior, we notify all dependent active expressions about the state change, thus, achieving the desired semantics.

### 4.4 Conceptual Limitations

Each of the three presented implementation strategies relies on a different underlying mechanism to capture state changes. These mechanisms imply certain limitations on the usage of active expressions that needs to be taken into consideration on which strategy to use for the task at hand.

**Convention**   The convention strategy (section 4.1) relies on the ability of the user to inject notifications manually into monitored source code, and can, in general, not guarantee the desired semantics of active expressions.

The convention strategy is closely related to the observer pattern [11]. As such, the convention strategy seems less convenient to use compared to the other two strategies and shares some common disadvantages with the observer pattern. For instance, if the programmer has no access to the source code to inject notifications at the appropriate points, the strategy is not applicable. Apart from that, the convention strategy represents a simple, library-based approach that uses only built-in language features. As a consequence, the strategy is not limited to a subset of language features but can deal with any kind of JavaScript expression.

**Interpretation**   The interpretation strategy (section 4.2) relies on two meta programming mechanisms, property accessors and the JavaScript-in-JavaScript interpreter, both coming with their own limitations:

- **Property Accessors**   Capturing state changes requires a means to intercept variable lookup to inject custom behavior. This requirement is only partially supported by JavaScript, namely for object fields. The usage of property accessors is limited to capture field storage and cannot intercept storage operations on local variables. Additionally, it is problematic to combine the interpretation strategy with other meta programming concepts that use property accessors as well, such as ContextJS [20]. ContextJS alternates the system behavior by modifying the very same property accessors used by the interpretation strategy. In such cases, we cannot guarantee proper semantics of active expressions as actual behavior depends on the order in which different property accessors are applied.

- **Interpreter** The underlying JavaScript-in-JavaScript interpreter provides only basic functionality. Naturally, the interpreter handles only a subset of ECMAScript3. Yet, one may use the babel compiler to mitigate this limitation by transpiling newer language features into legacy-compliant code. We further modified the interpreter to perform nested function interpretation and property lookup along the prototype chain. When interpreting nested function calls, the functions must not access local scope, otherwise they cannot be interpreted without access to an explicit scope object. The same holds true for functions that are passed as dynamic argument to the active expression construction. Furthermore, nested interpretation stops at native functions as no source code to interpret is available for these functions. In this case, the interpretation cannot be resumed when the native function calls a user function.

**Compilation** The compilation strategy (section 4.3) employs a source code transformation via a babel plugin that has to be integrated into the build tool chain of a project. Only transformed files are taken into account for state monitoring. Thus, the compilation strategy cannot capture state in native code as it cannot be rewritten. Similarly, we do not rewrite code dynamically executed using the eval function. However, transformed code interfaces seamlessly with non-transformed code. The compilation strategy captures assignments to local and global variables as well as object fields and is able to deal with expressions provided as dynamic arguments. Another downside of the compilation strategy is its relatively high performance overhead, as we describe in the following section.

### 4.5 Performance Analysis

The question which implementation strategy to use is not only affected by conceptual limitations but is also driven by the runtime overhead imposed by the strategies. Thus, we identify and compare the performance penalties implied by the different implementation strategies. We describe the micro benchmark in appendix B and discuss the main outcome in the following.

**Construction of Active Expressions** We first compare the time it takes to create an active expression in the three implementation strategies. As described in section B.2, the convention strategy is the fastest of the three strategies, because the system only adds the new active expression to a global set. Both other strategies need to determine the correct dependencies of the new active expression. The interpretation strategy is more than two orders of magnitude slower than the compilation strategy, because it creates a full-fledged JavaScript-in-JavaScript interpreter before performing the actual dependency analysis. The concrete overhead compared to the convention strategy is highly subjective to the complexity of the given expression.

**State Change Detection** Next, we identify the overhead introduced by the change detection mechanisms of the implementation strategies. In section B.3, we compare their performance against a baseline implementation that directly applies the intended

**Active Expressions**

reactive behavior. Again, the convention strategy is the fastest of the three strategies and has only a relative slowdown of 1.15 compared to the baseline implementation. The interpretation strategy has a slowdown of 1.54 and the compilation strategy is nearly four times slower than the baseline implementation.

**Impact of Source Code Transformation**   The compilation strategy imposes a slowdown even if no active expression is used. The slowdown is especially high for data intensive computations, as shown in section B.4. The reason for this high overhead is that the compilation strategy transforms all accesses to variables and object properties into function calls. These calls are *highly polymorphic* and, thus, hard to optimize by JITs.

**Interpretation vs. Compilation**   Finally, we analyze how the interpretation and the compilation strategy compare for varying numbers of active expressions in section B.5. As mentioned above, the compilation strategy has a high initial cost. As a result, the compilation strategy is nearly two orders of magnitude slower than the interpretation strategy if no active expressions are used. When increasing the number of active expressions in the system, the compilation strategy closes up to the interpretation strategy up to a point, where none is significantly faster. When increasing the number of active expressions even more, the relative slowdown of the compilation strategy increases again. This is due to the fact that the compilation strategy uses a centralized update mechanism, while the interpretation strategy associates each property accessor with the corresponding reactive behavior independently.

To summarize our findings, the convention strategy introduces the least overhead of all three implementation strategies for both, the construction of active expressions and the detection of changes. The interpretation strategy introduces a very high initial overhead for creating active expressions and a moderate overhead for change detection. In contrast, the compilation strategy implies performance penalties for the overall program, not just active expression-related parts.

## 4.6  Applicability

No single implementation strategy outperforms all others in all possible cases. Instead, each strategy has unique properties and is suitable for specific use cases. Based on the analyses above, we identify certain indicators of when and where to apply the different strategies:

The convention strategy provides fine-granular control over when and which active expressions are updated, but usually requires more manual work by the user. This manual overhead becomes more manageable for application domains in which the exact time of a state change is not crucial. Examples for this kind of domains are those that involve a global loop architecture, such as games, simulations, and graphical user interfaces. In such domains, the notification mechanism can simply be called once per frame for many practical use cases. The convention strategy works well in such domains as it fits into the overall architecture. However, if you require the exact semantics of active expressions, other strategies are more appropriate.

The interpretation strategy provides a fast state change detection mechanism and only monitors necessary portions of the system state. Thus, this strategy limits performance penalties even if active expressions are scattered around the entire program. However, the usage of a JavaScript-in-JavaScript interpreter implies several limitations, including the required explicit scope for each interpreted function, the inability to interpret native code, as well as the complicated interaction with certain meta programming concepts. Additionally, its high initial overhead suggests using the interpretation strategy when working with long-living active expressions.

The compilation strategy is able to deal with many use cases the other strategies cannot. However, the compilation strategy requires commitment to this particular strategy, because its source code transformation has a performance impact on all transformed modules, even if active expressions are not used. A strong indicator to use compilation over the other strategies is that active expressions are limited to some modules only. In this case, one may transform only said modules and, thereby, limit the overall overhead of the transformation.

## 5 Using Active Expressions to Implement State-based Reactive Concepts

This section outlines how to use active expressions to implement state-based reactive programming concepts. In particular, we exemplify the implementation of the four concepts described in our initial motivation in section 2: signals, constraints, reactive object queries, and implicit layer activation.[6] For each concept, we provide an implementation using active expressions as described in section 4 as well as a reference implementation without active expressions.[7] In the following, we describe and compare both implementations for reactive object queries. We present the implementations for the three remaining concepts in appendix C. Finally, we compare the active expression-based implementations against the reference implementations in terms of code complexity.

### 5.1 Reactive Object Queries

As described in section 2, reactive object queries allow to query the program space for all instances of a class that fulfill a given condition. The set of instances updates automatically whenever the program state changes. As a result, the set maintains consistency with the program state. In addition to querying objects, one may apply typical collection operations on these sets to refine and process them. Refined sets update the same way queried sets do. We compare an implementation based on active expressions with an implementation based on the original work [19]. However, for

---

[6] We simplify details that are not relevant in the context of this work.
[7] All implementations are available in their respective repositories in https://github.com/active-expressions accessed on November 30th 2016.

**Active Expressions**

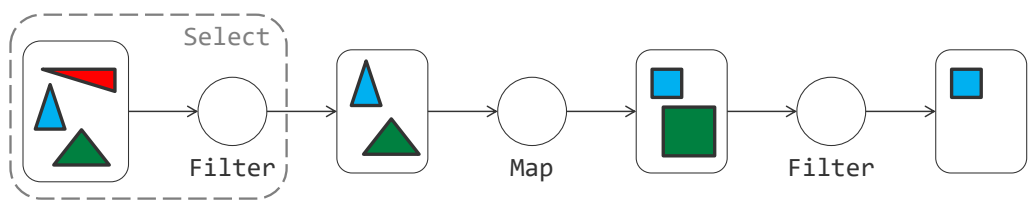

**Figure 6** A graph created with **select**(Triangle, isIsosceles).map(toSquares).filter(isBlue);

better comparison, we provide an adjusted implementation[8] that uses the same key concepts as the original work, but has access to the same JavaScript features available to our active expression-based implementation.

**Change Detection for Reactive Object Queries**  As illustrated in figure 6, reactive object queries are organized as a bipartite graph of sets and operators. Each set but the base set is created and maintained by its respective operator. The base set is populated with all instances of a class using a functional mixin to install an after advice on the respective initialize method. When querying for objects using **select**(Class, condition), the framework actually applies a FilterOperation to the corresponding base set:

```
1  export default function select(Class, expression, context) {
2    let newSelection = new View();
3    new FilterOperator(Class._instances_, newSelection, expression, context);
4    return newSelection;
5  }
```

The FilterOperation created in line 3 is responsible for maintaining the resulting set according to the given expression. To do so, whenever a new instance is created, we identify the dependencies needed for change detection using an ExpressionObserver:

```
1  class FilterOperator extends IdentityOperator {
2    onNewInstance(item, context) {
3      new ExpressionObserver(
4        this.expression,
5        context,
6        item,
7        () => this.conditionChanged(item)
8      );
9
10     if(this.expression(item)) {
11       this.add(item);
12     }
13   }
14 }
```

We provide the conditionChanged method as callback for changes in line 7. In reaction to a change, this method is updates the depending set by adding the instance to or remove it from the set accordingly. Additionally, we initialize the correct initial state for the instance in line 10 to 12.

---

[8] https://github.com/active-expressions/programming-roq-plain accessed on February 19th 2016, at commit ocad577

Similar to the interpretation strategy described in section 4.2, the `ExpressionObserver` uses dynamic interpretation to identify dependencies. A stack of `ExpressionObservers` ensures that dependencies are correctly set up in case of nested interpretations:

```
1  export class ExpressionObserver {
2    installListeners() {
3      // in case of multiple interpretations on top of each other
4      expressionObserverStack.withElement(this, () => {
5        ExpressionInterpreter.runAndReturn(this.func, this.scope, this.item);
6      });
7    }
8  }
```

Whenever the customized JavaScript-in-JavaScript interpreter visits a property access, we introduce a dependency between this property access and the current `ExpressionObserver`:

```
1  export class ExpressionInterpreter extends Interpreter {
2      getProperty(obj, name) {
3          let object = obj.valueOf(),
4              prop = name.valueOf();
5  
6          PropertyListener
7              .watchProperty(object, prop)
8              .addHandler(expressionObserverStack.top());
9  
10         return super.getProperty(obj, name);
11     }
12 }
```

When the class `PropertyListener` is asked to watch a property in line 7, it either creates a new listener instance or reuses an existing one if existent for the particular object property. The `PropertyListener` mediates between `PropertyAccessors` and any number of observers, so that a change detected by a `PropertyAccessor` is propagated to each depending observer. Furthermore, the listener handles re-interpretation in case of a structural change in the dependencies.

The class `PropertyAccessor` is responsible for actual state change detection and, thus, interfaces with native property accessors. As a consequence, the class has to deal with a lot of low-level responsibilities, for example, wrapping existing property accessors if existent in line 3 and dealing with browser-specific behavior to properly handle `Array.length` in line 5 to 11:

**Active Expressions**

```
1  export class PropertyAccessor {
2    constructor(obj, propName) {
3      this.safeOldAccessors(obj, propName);
4
5      try {
6        obj.__defineGetter__(propName, () => this[PROPERTY_ACCESSOR_NAME]);
7      } catch (e) { /* Firefox raises for Array.length */ }
8      let newGetter = obj.__lookupGetter__(propName);
9      if (!newGetter) { // Chrome silently ignores __defineGetter__ for Array.length
10       return;
11     }
12     // [...]
13   }
14 }
```

**Active Expression-based Implementation** We refactor the existing implementation of reactive object queries to use active expressions.[9] Thus, we need to provide the method onNewInstance for the class FilterOperation to implement its reactive semantics:

```
1  class FilterOperator extends IdentityOperator {
2    onNewInstance(item) {
3      trigger(aexpr(() => this.expression(item)))
4        .onBecomeTrue(() => this.add(item))
5        .onBecomeFalse(() => this.remove(item));
6    }
7  }
```

The method above is responsible for adding the instance to or remove it from the set according to its changing state. To do so, we create an active expression to monitor the expression parametrized with the given instance in line 3. For the reaction to a change, we make use of *triggers*, a small utility library build on top of active expressions. The full source code of triggers is provided in appendix D. The trigger function takes an active expression and provides a set of methods to reason about changes to the expression result. For example, callbacks given to the onBecomeTrue method are executed, whenever the result of the expression becomes true (from a non-true value) and initially, if the result of the expression is currently true. In particular, we add the given instance to the set in line 4, whenever the expression result becomes true, or if the expression evaluates to true initially. Analogously, we remove the instance from the view in line 5, whenever the expression result becomes false from a non-false value.

**Discussion** In its original variant, the concept implementor has to implement a vertical slice through multiple layers of abstractions, from the high-level view of expressions to browser-specific behavior regarding their native property accessor implementation. We used multiple classes to bridge the gap between a high-level

---

[9] https://github.com/active-expressions/reactive-object-queries accessed on February 18th 2016, at commit f45fe5e

■ **Table 1** Code complexity of the presented examples in terms of the number of AST nodes.

| AST nodes | plain | active expression |
|---|---|---|
| Signals | 1707 | 274 |
| Linear Constraints | 1267 | 672 |
| Reactive Object Queries | 2820 | 2092 |
| Implicit Layer Activation | 72 | 40 |

abstraction and its most fine-granular implementation details. With our concept, we were able to replace this complex, state-monitoring code dealing with reflection and dynamic interpretation with a single active expression. Furthermore, the presented implementation exemplifies that active expressions can integrate with existing reactive solutions.

## 5.2 Impact

We were able to implement the presented four state-based reactive concepts using active expressions to detect and react to state changes, without having to deal with tedious implementation details, such as meta programming, reflection, static code analysis, and dynamic interpretation. To measure the impact of active expressions, we quantitatively compare the active expression-based implementations with their respective reference implementations in terms of code complexity. As a measurement for code complexity, we count the total number of AST nodes in an implementation, as described in appendix E. We summarize our measurements in table 1. According to table 1, active expressions reduce the code complexity for implementing state-based reactive concepts compared to other meta programming techniques in JavaScript. Important to note, active expressions only target the detection of changes, not the reaction to those changes. Thus, the code required to describe reactive behavior sets a natural lower bound for each of the presented implementations. However, as the detection part typically involves unnecessary complex implementation details, programmers may focus more on the reaction of their implementation when using active expressions. As a result, we conduct active expressions to be a useful primitive regarding the implementation of state-based reactive concepts.

## 6 Related Work

Active expressions are no constructed programming concept, but rather a natural conclusion of our observations on existing reactive programming approaches. As such, active expressions act as a possible foundation for state-based reactive concepts, including incremental lists, two-way data-bindings, and the concepts presented in section 2. This section presents fields of research and programming approaches related to active expressions.

**Active Expressions**

**Event-based Programming**  Event-based programming systems [7, 12] rely on explicit signaling of occurrences in the execution of a program as named events [29]. Active expressions relate to event-based programming in that they lift modifications of the program state onto the level of events. This lifting enables user code to access these otherwise hidden state modifications in the execution and react on them. With complex modifications of state being available as events, they become subject of the ordinary event handling. An integration of active expressions with an event-based environment could therefore handle both, complex state modifications and named events, in a unified manner [24].

**Aspect-oriented Programming**  Both, active expressions and aspect-oriented programming (AOP) [17], turn certain actions during the execution of a program into explicit events. In contrast to active expressions, aspect-oriented programming (AOP) reifies a number of different actions, including function calls and object instantiations, as events. This reification allows AOP users to react on, enhance, or even replace said actions. AOP represents another possible implementation strategy for active expressions. To do so, some meta programming capabilities are required from the underlying language: to implement active expressions, the AOP framework has to be able to intercept all kinds of state-modifying actions, such as assignments to local variables or object members, as pointcuts. Analogous, active expressions can be used to implement AOP pointcuts. However, to implement a specific pointcut using active expressions, the underlying execution environment needs to make the desired pointcut accessible in form of state.

**Object Constraint Programming**  In object constraint programming, one specifies relations that should hold as expressions over state. To detect invalidations of these constraint expressions, the Babelsberg family [8] of object constraint programming languages uses concepts similar the interpretation strategy (section 4.2), namely property accessors [9] and method wrappers [13]. Thus, Babelsberg has similar detection capabilities as active expressions. However, Babelsberg uses a fixed reactive behavior dedicated to repair invalidated constraints. This complex solving behavior is intentionally hidden from its users in cooperating solvers. In contrast to Babelsberg that aims for expressiveness using specific but narrow reactive behavior, active expressions provide generality by incorporating arbitrary user-defined reactions.

**Explicit Lifting**  Explicit lifting represents another possible strategy for state change detection [3]. Similar to the convention strategy described in section 4.1, explicit lifting imposes restrictions to the user of a concept. In particular, systems such as the Flapjax library [26] and Frappé [5] require primitive operations to be lifted on reactive mechanisms such as behaviors, so that these mechanisms may explicitly notify dependent components about change. As a consequence, all code that should be able to notify dependent components needs to be adapted. However, one of our requirements is to integrate with already existing imperative libraries, as described in section 3. Thus, we prefer implicit lifting as implemented by the compilation strategy (section 4.3) over explicit lifting.

**Wallingford**   The constraint reactive programming language Wallingford is capable to track changing values over time [4]. Most notably, Wallingford's when constraints provide a semantics similar to active expressions: these constraints wait for a condition to become true and execute provided certain behavior in response. The major difference between the two systems regards the notion of time. Wallingford is a holistic system to express time-continuous simulations. In contrast, active expressions deal with changing variable values in discrete-time. Their discrete notion of time eases the integration of active expressions with existing OOP environments.

**Watchers**   Watchers [33] are a debugging technique to observe the runtime state of variables or expressions in the code. Active expressions make this debugging technique available within a program itself. As a result, application code can install watchers and react to changes in the observed system state.

## 7  Conclusion and Future Work

*Active expressions* address the question of how to *ease the development* of novel reactive programming concepts. We approach this question by identifying and exploiting *commonalities* in existing reactive concepts. In particular, we identify a subset of reactive programming concepts that share a common change detection mechanism: each *state-based reactive concept* reacts to changes in the evaluation result of a given expression. Implementing detection mechanisms represents a recurring necessity when developing practical reactive programming concepts. Thus, we design active expressions as a reusable primitive for state monitoring to ease the implementation of state-based reactive concepts. We provide an implementation of active expressions with variable state monitoring strategies. Furthermore, we provide examples to demonstrate the applicability of active expressions to a number of state-based reactive concepts. These examples indicate that one can implement reactive concepts using active expressions without having to care about tedious implementation details of change detection. As a result, active expressions lower the entrance barrier to the development of reactive programming systems. We hope that active expressions help to encourage more people to experiment with novel reactive concepts and, thus, advance our field of research.

We expect to continue to evolve both the concept of active expressions and its implementation:

**Active Expressions as a Language Primitive**   We designed active expressions as a common foundation for state-based reactive concepts. The provided examples hint the potential of active expressions. One direction for future work is to broaden the scope of active expressions: instead of using active expressions as a solution specific to state-based reactive concepts, we could make active expressions a language primitive. Having this new primitive may open up interesting opportunities. What patterns will we discover in such a language? And, do these patterns differ from the ones we observe in reactive concepts that are build on top of existing languages?

**Active Expressions**

Providing active expressions as a language primitive involves two major tasks. First, instead of treating active expressions specially, every expression in the language should be active. Second, the declaration of dependencies between expressions and behavior should be integrated syntactically into the language.

**Improve Comprehensibility through Tool Support**   Similar to other high-level concepts, active expressions encapsulate complex behavior. Thus, even clean and simple code might be hard to comprehend in context of a larger application without the appropriate tool support. To make active expressions practically usable, tools should support the code comprehension in both, static and dynamic settings. For example, static analysis tools might identify which statements might trigger an active expression. A dedicated debugging tool is especially important, because most debuggers have been designed with imperative concepts in mind and, thus, are unsuited for reactive programming concepts [30]. In particular, a dedicated debugging tool might reveal what active expressions exist in the system, how they relate to each other, and what state they depend on.

**Acknowledgements**   We gratefully acknowledge the financial support of the Research School of the Hasso Plattner Institute. We thank Patrick Rein and Jens Lincke for discussions on earlier versions of this submission.

**References**

[1]   Malte Appeltauer, Robert Hirschfeld, Michael Haupt, Jens Lincke, and Michael Perscheid. "A Comparison of Context-oriented Programming Languages". In: *International Workshop on Context-Oriented Programming (COP)*. Genova, Italy: ACM, 2009, 6:1–6:6. ISBN: 978-1-60558-538-3. DOI: 10.1145/1562112.1562118.

[2]   Greg J. Badros, Alan Borning, and Peter J. Stuckey. "The Cassowary Linear Arithmetic Constraint Solving Algorithm". In: *ACM Transactions on Computer-Human Interaction (TOCHI)* 8.4 (Dec. 2001), pages 267–306. ISSN: 1073-0516. DOI: 10.1145/504704.504705.

[3]   Engineer Bainomugisha, Andoni Lombide Carreton, Tom van Cutsem, Stijn Mostinckx, and Wolfgang de Meuter. "A Survey on Reactive Programming". In: *ACM Computing Surveys (CSUR)* 45.4 (Aug. 2013), 52:1–52:34. ISSN: 0360-0300. DOI: 10.1145/2501654.2501666.

[4]   Alan Borning. "Wallingford: Toward a Constraint Reactive Programming Language". In: *Constrained and Reactive Objects Workshop (CROW)*. MODULARITY Companion 2016. Málaga, Spain: ACM, 2016, pages 45–49. ISBN: 978-1-4503-4033-5. DOI: 10.1145/2892664.2892667.

[5]   Antony Courtney. "Frappé: Functional Reactive Programming in Java". In: *Third International Symposium on Practical Aspects of Declarative Languages (PADL) March 11–12*. Las Vegas, Nevada, USA: Springer, 2001, pages 29–44. ISBN: 978-3-540-45241-6. DOI: 10.1007/3-540-45241-9_3.

[6]  Conal Elliott and Paul Hudak. "Functional Reactive Animation". In: *Second ACM SIGPLAN International Conference on Functional Programming (ICFP)*. Amsterdam, The Netherlands: ACM, 1997, pages 263–273. ISBN: 0-89791-918-1. DOI: 10.1145/258948.258973.

[7]  Patrick Eugster and K. R. Jayaram. "EventJava: An Extension of Java for Event Correlation". In: *23rd European Conference on Object-Oriented Programming (ECOOP), July 6-10, 2009*. Genoa, Italy: Springer, 2009, pages 570–594. ISBN: 978-3-642-03013-0. DOI: 10.1007/978-3-642-03013-0_26.

[8]  Tim Felgentreff, Alan Borning, and Robert Hirschfeld. "Specifying and Solving Constraints on Object Behavior". In: *Journal of Object Technology (JOT)* 13.4 (Sept. 2014), 1:1–38. ISSN: 1660-1769. DOI: 10.5381/jot.2014.13.4.a1.

[9]  Tim Felgentreff, Alan Borning, Robert Hirschfeld, Jens Lincke, Yoshiki Ohshima, Bert Freudenberg, and Robert Krahn. "Babelsberg/JS — A Browser-Based Implementation of an Object Constraint Language". In: *28th European Conference on Object-Oriented Programming (ECOOP), July 28 – August 1, 2014*. Uppsala, Sweden: Springer, 2014, pages 411–436. ISBN: 978-3-662-44202-9. DOI: 10.1007/978-3-662-44202-9_17.

[10]  Bjørn N. Freeman-Benson. "Kaleidoscope: Mixing Objects, Constraints and Imperative Programming". In: *Conference on Object-Oriented Programming Systems, Languages, and Applications (OOPSLA)*. Ottawa, Canada: ACM, 1990, pages 77–88. ISBN: 0-89791-411-2. DOI: 10.1145/97945.97957.

[11]  Erich Gamma, Richard Helm, Ralph Johnson, and John Vlissides. *Design patterns: elements of reusable object-oriented software*. Addison-Wesley, 1995. ISBN: 0-201-63361-2.

[12]  Vaidas Gasiunas, Lucas Satabin, Mira Mezini, Angel Núñez, and Jacques Noyé. "EScala: modular event-driven object interactions in scala". In: *10th International Conference on Aspect-Oriented Software Development (AOSD), March 21-25, 2011*. Porto de Galinhas, Brazil: ACM, 2011, pages 227–240. ISBN: 978-1-4503-0605-8. DOI: 10.1145/1960275.1960303.

[13]  Maria Graber, Tim Felgentreff, Robert Hirschfeld, and Alan Borning. "Solving Interactive Logic Puzzles With Object-Constraints — An Experience Report Using Babelsberg/S for Squeak/Smalltalk". In: *Workshop on Reactive and Event-based Languages & Systems (REBLS)*. 2014, 1:1–1:5.

[14]  Martin Grabmüller and Petra Hofstedt. "Turtle: A Constraint Imperative Programming Language". In: *Research and Development in Intelligent Systems XX: Proceedings of AI2003, the 23rd SGAI International Conference on Innovative Techniques and Applications of Artificial Intelligence*. Springer, 2004, pages 185–198. ISBN: 978-0-85729-412-8. DOI: 10.1007/978-0-85729-412-8_14.

[15]  Robert Hirschfeld, Pascal Costanza, and Oscar Nierstrasz. "Context-oriented Programming". In: *Journal of Object Technology (JOT)* 7.3 (Mar. 2008), pages 125–151. ISSN: 1660-1769. DOI: 10.5381/jot.2008.7.3.a4.

**Active Expressions**

[16] Tetsuo Kamina, Tomoyuki Aotani, and Hidehiko Masuhara. "Generalized layer activation mechanism through contexts and subscribers". In: *14th International Conference on Modularity (MODULARITY), 2015*. Fort Collins, Colorado, USA: ACM, 2015, pages 14–28. ISBN: 978-1-4503-3249-1. DOI: 10.1145/2724525.2724570.

[17] Gregor Kiczales, John Lamping, Anurag Mendhekar, Chris Maeda, Cristina Lopes, Jean-Marc Loingtier, and John Irwin. "Aspect-oriented programming". In: *11th European Conference on Object-Oriented Programming (ECOOP), June 9–13, 1997*. Jyväskylä, Finland: Springer, 1997, pages 220–242. ISBN: 978-3-540-69127-3. DOI: 10.1007/BFb0053381.

[18] Stefan Lehmann, Tim Felgentreff, and Robert Hirschfeld. "Connecting Object Constraints with Context-oriented Programming: Scoping Constraints with Layers and Activating Layers with Constraints". In: *7th International Workshop on Context-Oriented Programming (COP)*. Prague, Czech Republic: ACM, 2015, 1:1–1:6. ISBN: 978-1-4503-3654-3. DOI: 10.1145/2786545.2786549.

[19] Stefan Lehmann, Tim Felgentreff, Jens Lincke, Patrick Rein, and Robert Hirschfeld. "Reactive Object Queries". In: *Constrained and Reactive Objects Workshop (CROW)*. MODULARITY Companion 2016. Málaga, Spain: ACM, 2016. ISBN: 978-1-4503-4033-5. DOI: 10.1145/2892664.2892665.

[20] Jens Lincke, Malte Appeltauer, Bastian Steinert, and Robert Hirschfeld. "An Open Implementation for Context-oriented Layer Composition in ContextJS". In: *Science of Computer Programming (SCICO)* 76.12 (2011), pages 1194–1209. ISSN: 0167-6423. DOI: 10.1016/j.scico.2010.11.013.

[21] Martin von Löwis, Marcus Denker, and Oscar Nierstrasz. "Context-oriented Programming: Beyond Layers". In: *International Conference on Dynamic Languages (ICDL), 2007*. Lugano, Switzerland: ACM, 2007, pages 143–156. ISBN: 978-1-60558-084-5. DOI: 10.1145/1352678.1352688.

[22] Ingo Maier and Martin Odersky. "Higher-order reactive programming with incremental lists". In: *27th European Conference on Object-Oriented Programming, (ECOOP), July 1-5, 2013*. Montpellier, France: Springer, 2013, pages 707–731. ISBN: 978-3-642-39038-8. DOI: 10.1007/978-3-642-39038-8_29.

[23] Erik Meijer. "Reactive extensions (Rx): curing your asynchronous programming blues". In: *ACM SIGPLAN Commercial Users of Functional Programming (CUFP)*. ACM. 2010, page 11. ISBN: 978-1-4503-0516-7.

[24] Erik Meijer, Brian Beckman, and Gavin M. Bierman. "LINQ: reconciling object, relations and XML in the .NET framework". In: *International Conference on Management of Data (SIGMOD), June 27-29, 2006*. Chicago, Illinois, USA: ACM, 2006, page 706. ISBN: 1-59593-434-0. DOI: 10.1145/1142473.1142552.

[25] Kim Mens, Rafael Capilla, Nicolás Cardozo, and Bruno Dumas. "A taxonomy of context-aware software variability approaches". In: *Workshop on Live Adaptation of Software SYstems (LASSY), March 14 - 18, 2016*. MODULARITY Companion 2016. Málaga, Spain: ACM, 2016, pages 119–124. ISBN: 978-1-4503-4033-5. DOI: 10.1145/2892664.2892684.

[26] Leo A. Meyerovich, Arjun Guha, Jacob P. Baskin, Gregory H. Cooper, Michael Greenberg, Aleks Bromfield, and Shriram Krishnamurthi. "Flapjax: A Programming Language for Ajax Applications". In: *24th ACM SIGPLAN Conference on Object-Oriented Programming Systems, Languages, and Applications (OOPSLA), 2009*. Orlando, Florida, USA: ACM, 2009, pages 1–20. ISBN: 978-1-60558-766-0. DOI: 10.1145/1640089.1640091.

[27] John F. Pane, Chotirat (Ann) Ratanamahatana, and Brad A. Myers. "Studying the language and structure in non-programmers' solutions to programming problems". In: *International Journal of Human-Computer Studies* 54.2 (2001), pages 237–264. ISSN: 1071-5819. DOI: 10.1006/ijhc.2000.0410.

[28] Guido Salvaneschi, Sven Amann, Sebastian Proksch, and Mira Mezini. "An empirical study on program comprehension with reactive programming". In: *22nd ACM SIGSOFT International Symposium on Foundations of Software Engineering (FSE), 2014*. ACM. Hong Kong, China: ACM, 2014, pages 564–575. ISBN: 978-1-4503-3056-5. DOI: 10.1145/2635868.2635895.

[29] Guido Salvaneschi, Patrick Eugster, and Mira Mezini. "Programming with Implicit Flows". In: *IEEE Software* 31.5 (Sept. 2014), pages 52–59. ISSN: 0740-7459. DOI: 10.1109/MS.2014.101.

[30] Guido Salvaneschi and Mira Mezini. "Debugging for reactive programming". In: *38th International Conference on Software Engineering (ICSE), May 14-22, 2016*. Austin, Texas, USA: ACM, 2016, pages 796–807. ISBN: 978-1-4503-3900-1. DOI: 10.1145/2884781.2884815.

[31] Christopher Schuster and Cormac Flanagan. "Reactive programming with reactive variables". In: *Constrained and Reactive Objects Workshop (CROW)*. MODULARITY Companion 2016. Málaga, Spain: ACM, 2016, pages 29–33. ISBN: 978-1-4503-4033-5. DOI: 10.1145/2892664.2892666.

[32] Tom Van Cutsem and Mark S Miller. "Proxies: design principles for robust object-oriented intercession APIs". In: *6th Symposium on Dynamic Languages (DLS), 2010*. Volume 45. 12. ACM. Reno/Tahoe, Nevada, USA: ACM, 2010, pages 59–72. ISBN: 978-1-4503-0405-4. DOI: 10.1145/1869631.1869638.

[33] Qin Zhao, Rodric Rabbah, Saman Amarasinghe, Larry Rudolph, and Weng-Fai Wong. "How to Do a Million Watchpoints: Efficient Debugging Using Dynamic Instrumentation". In: *17th International Conference on Compiler Construction (CC), Held as Part of the Joint European Conferences on Theory and Practice of Software (ETAPS), March 29 - April 6, 2008*. Budapest, Hungary: Springer, 2008, pages 147–162. ISBN: 978-3-540-78791-4. DOI: 10.1007/978-3-540-78791-4_10.

## A  Usage of Babel Quantified

To analyze the portion of JavaScript projects that make use of babel, we use GHTorrent, a project to allow large-scale analyses on Github data.[10] We use a data set containing Github data until December 2016, inclusive. In total, the data set lists 10,590,075 projects on Github that contain any code classified as JavaScript. For further analysis, we only consider projects that fulfill the following criteria:

1. The project is mainly classified as a JavaScript project (language='JavaScript').
2. The project was created recently, i.e. in 2015 or 2016 (**substring**(created_at, 1, 4) = '2015').
3. The project is not a fork of another project (forked_from is **null**). We added this criteria to exclude projects created in context of bootcamps or tutorials.

With these criteria, we identify the overall number of JavaScript projects per year, 2016 in the following example:

```
select count(*) from projects where language='JavaScript' and substring(created_at, 1, 4) = '2016'
    ↪ and forked_from is null;
```

The results contain 849,662 projects in 2015 and 395,319 projects in 2016. Next, we identify the number of projects that use babel as a dependency. To do so, we further assume the projects to use npm as package manager. Thus, we check the package.json for the respective dependency. As GHTorrent does not provide full source code but only patches, we identify all commits that modified the package.json. Then, we check for patches that contain the substring 'babel'. This includes several projects one may depend on to use babel, such as babel-cli, babel-core, babel-standalone, and babelify:

```
select t.project_id, t.project_name, t.user_login, count(fp.id) as forks from temp_babel_projects
    ↪ as t, (projects as p left outer join projects as fp on (fp.forked_from = p.id)) where t.
    ↪ project_id = p.id and p.language = 'JavaScript' and substring(p.created_at, 1, 4) = '2016'
    ↪ and p.forked_from is null group by t.project_id, t.project_name, t.user_login order by
    ↪ forks desc;
```

Overall, we identified 1,281,048 commits that belong to 210,266 projects. Finally, we join both, the JavaScript projects and the projects that modified their package.json regarding babel:

```
select t.user_login, t.project_name, count(pf.id) as forks from temp_babel_projects as t,
    ↪ projects as p left outer join projects as pf on (pf.forked_from = p.id) where t.project_id
    ↪ = p.id group by t.user_login, t.project_name order by forks desc;
```

We summarize our findings in table 2. Most notably, over 10% percent of all JavaScript projects created in 2016 depend on babel.

**Treads of Validity**  There are a number of points to be taken into consideration regarding this analysis:

- Our data set is restricted to Github repositories.
- We only considered projects that use npm for package management. This excludes projects packaged using other means like bower or that use no package manager

---

[10] http://ghtorrent.org/ accessed on February 27th 2017

■ **Table 2** Results of a quantitative analysis of JavaScript projects on Github regarding the usage of babel

| Year created | Total JavaScript Projects | Using Babel | Portion in percent |
|---|---|---|---|
| 2015 | 849,662 | 56,998 | 6.71 |
| 2016 | 395,319 | 49,719 | 12.58 |

at all. Also, this overview does not contain projects of users that installed and use babel globally on their system.
- We search for the string 'babel' in patches of the package.json. This may return some false positives as the string could be part of another dependency or does not appear in the dependency section of the package.json.
- We could only check patches. Thus, our analysis includes projects that used babel for a certain amount of time and then removed it.

## B  Micro Benchmarks

To identify the performance penalties implied by the different implementation strategies described in section 4.1 to 4.3, we provide and discuss multiple benchmark scenarios in the following. We summarize the results in section 4.5.

### B.1  Performance Benchmark Setup and Statistical Methods

**Technical Specification**   All benchmarks were executed on the following system:
- CPU and memory: Intel(R) Core(TM) i7-6650U CPU @ 2.20GHz 2.21 GHz, 4 Logical cores; 16.0 GB Main Memory
- System software: Windows 10 Pro, version 1511 (OS Build 10586.545)
- Runtime: Google Chrome version 52.0.2743.116 m; benchmarks executed using Karma test runner version 1.2.0 and Mocha test framework version 3.0.2
- Transpiler and bundler: babel-cli 6.11.4 (no es2015 preset) and rollup 0.34.8,
- Libraries under test: active-expressions 1.5.0, aexpr-source-transformation-propagation 1.4.2, babel-plugin-aexpr-source-transformation 2.2.0
- Benchmark suite: active-expressions-benchmark at commit c37d44d.[11]

We measured the execution time of a benchmark by wrapping the benchmark in a function and measuring the time between calling the function and it returning. Each benchmark configuration was iterated 100 times, with only the final 30 iterations taken into account for the overall performance measurement to mitigate the effects of the V8 JIT.

---

[11] https://github.com/active-expressions/active-expressions-benchmark accessed on September 30th 2016

**Active Expressions**

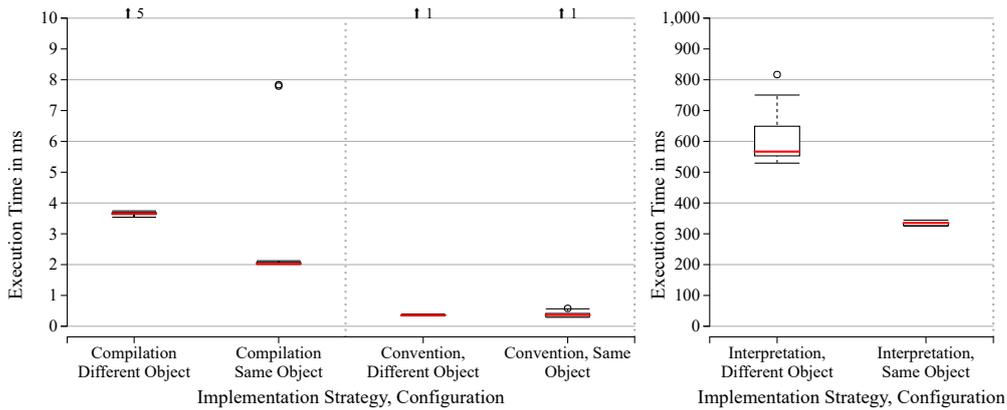

**(a)** Execution times for Convention and Compilation

**(b)** Execution times for Interpretation

■ **Figure 7** Execution times for the construction of 1000 active expressions monitoring the aspect ratio of a rectangle. The red line indicates the median, the upper and lower edges of the box are the second and third quartile and the end of the whiskers are the most outlying values in a 1.5 inter-quartile range distance from the second and third quartile. Arrows with a number above a boxplot indicate that there are that many outliers out of the chart scale. Note, that figure 7b is scaled up by two magnitudes compared to figure 7a.

**Statistical Methods** We make no assumptions on the underlying distribution and provide Tukey boxplots in figure 7 to 10 to visualize the median and variation of the measured timings. Exact median timings are given in table 3 to 6. Slowdowns are computed by dividing the median execution times of the measurements to compare. Confidence bounds of this statistic are given by the 2.5-th and 97.5-th percentile of the bootstrap distribution of the computed ratio.

**B.2 Construction of Active Expressions**

**Setup** First off, we analyze the initial cost to create active expressions for each implementation strategy. Therefore, we measure the time to create 1000 active expressions monitoring the aspect ratio of a rectangle. We distinguish two different cases: creating 1000 active expressions monitoring the same object and creating 1000 active expressions, each monitoring a different rectangle object.

**Discussion** As figure 7 reveals, the convention strategy has the lowest runtime for the construction of active expressions. This result is to be expected, as the convention strategy only adds the given expression to a global set when creating an active expression. In contrast, the compilation and interpretation strategies have to set up their respective dependency mechanisms during active expression construction. Accordingly, they impose high overhead as shown by the relative slowdowns in table 3. While the compilation strategy runs the given expression in native JavaScript, the interpretation strategy uses a full-fledged JavaScript-in-JavaScript interpreter to

■ **Table 3** Benchmark timings and relative slowdowns for creating 1000 active expressions on the aspect ratio of a rectangle. Slowdowns given as ratio of medians with 95% confidence intervals.

|  | timing [ms] | | slowdown |
| --- | --- | --- | --- |
|  | same object | diff. objects | diff. vs same |
| Convention | 0.37 | 0.36 | 0.98 [0.87 - 1.13] |
| Interpretation | 335.43 | 566.79 | 1.69 [1.66 - 1.77] |
| Compilation | 2.02 | 3.66 | 1.80 [1.80 - 1.82] |
|  | slowdown (vs Convention) | | |
|  | same object | diff. objects | |
| Interpretation | 912.73 [1032.09 - 993.84] | 1574.41 [1550.04 - 1707.67] | |
| Compilation | 5.51 [4.99 - 6.43] | 10.15 [10.10 - 10.76] | |

determine relevant dependencies, which explains the high impact of the interpretation strategy. The relative overhead compared to the convention strategy is subject to the complexity of the given expression.

Whether we monitor the same object or different objects influences the overhead for the interpretation and compilation strategy, although not as significant as the choice of the overall implementation strategy itself. As shown in table 3, the speedup of monitoring the same object in contrast to monitoring different objects is below two for interpretation and compilation. The interpretation strategy uses property accessors to recognize changes. When visiting the same object-property-combination, we reuse the existing property accessor instead of creating a new one, which explains the lower runtime when wrapping the same object. Similarly, the compilation strategy is able to reuse dependencies if already existent.

**B.3 State Change Detection**

**Setup** The above measurements show only the initial overhead of the respective implementation strategy. In the following, we want to identify the overhead that is imposed by actually using active expressions. Continuing the previous scenario, we monitor the aspect ratio of a given rectangle using an active expression. In an associated callback we adjust the height so that the aspect ratio matches a target value. We measure the time it takes to assign 100,000 randomly generated widths and test whether the aspect ratio is the desired value after each assignment. For the convention strategy we need to insert a point to check for state changes explicitly between the assignment and the test. For better comparison we provide a baseline approach that directly adjusts the height after the assignment and before the test to match the desired aspect ratio.

**Discussion** According to the results in figure 8 and table 4, the convention strategy has the lowest overhead among the three strategies when updating monitored results

**Active Expressions**

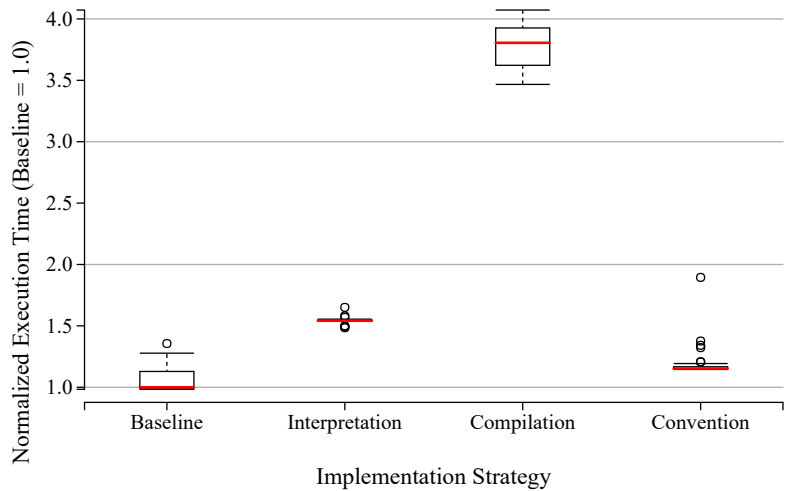

▪ **Figure 8** Performance benchmark results for the update and monitoring overhead of active expressions, normalized by baseline median. The normalization is the quotient of execution time of the respective implementation and the time of the baseline solution.

▪ **Table 4** Benchmark timings and relative slowdowns for updating the width of a rectangle with its aspect ratio monitored.

|                | timing [ms] | slowdown (vs Baseline) |
|----------------|-------------|------------------------|
| Baseline       | 57.17       |                        |
| Convention     | 65.87       | 1.15 [1.08 – 1.17]     |
| Interpretation | 88.13       | 1.54 [1.45 – 1.56]     |
| Compilation    | 217.54      | 3.81 [3.54 – 3.89]     |

frequently. However, note that in this benchmark, a high percentage of operations modify important system state. As a result, the interpretation and the compilation strategies resolve active expression callbacks very frequently. Considering the high overhead for the creation of active expressions and the moderate impact during updates indicates that the interpretation strategy pays off for long-living active expressions.

### B.4 Impact of Source Code Transformation

**Setup** The previous experiment hints a high overhead of the compilation strategy. Multiple mechanisms within this implementation strategy could potentially contribute to this high runtime overhead. To further examine this overhead, we want to examine the performance penalty introduced by the used source code transformation. The source code transformation replaces read and write accesses with proper function invocations, which potentially slow down the execution. So, as a worst case scenario, we choose an algorithm that primarily reads and writes data: performing quicksort on an array of 10000 randomly generated numbers. We determine the runtime of both, a plain, unmodified baseline version and the same code, but transformed. The transformation introduces detection hooks to system state changes. However, in this benchmark we do not create any active expressions to use these hooks, effectively eliminating the computational overhead of the callback notification mechanism.

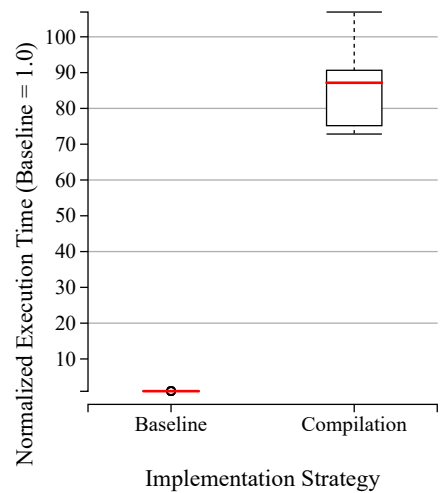

**Figure 9** Performance impact of the source code transformation for sorting a 10000 element array, normalized by the median for the non-transformed code execution.

**Table 5** Benchmark timings and relative slowdowns for sorting a 10000 element array.

|  | timing [ms] | slowdown (vs Baseline) |
|---|---|---|
| Baseline | 0.86 |  |
| Compilation | 74.96 | 87.16 [75.86 – 90.22] |

**Discussion**  As figure 9 reveals, the source code transformation imposes a high performance overhead, even when no active expression is used. The reason for this high overhead is that the invocation of detection hooks, such as `getMember`, is *highly polymorphic*, and, therefore hard to optimize by JITs. As every access to a property and every call of a member function is wrapped, this strategy can cause severe performance penalties. This issue becomes especially apparent for data intense problems that are otherwise easy to optimize. This explains why the compilation strategy is over 80 times slower that the baseline implementation in this scenario (table 5), but only about four times slower in the previous one (table 4).

### B.5 Interpretation vs. Compilation

**Setup**  Interpretation and compilation share similar mechanisms, compilation however has a big initial cost. We want to analyze how they relate to each other performance-wise for different numbers of active expressions. We use quicksort as an example again, with an array of 1000 randomly generated numbers. In each benchmark, we

**Active Expressions**

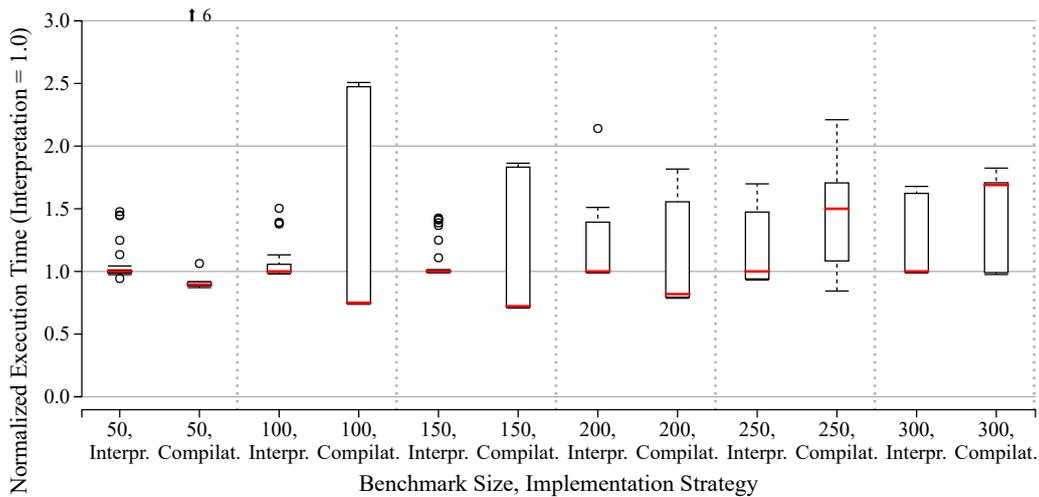

**Figure 10** Performance measurements of a quicksort implementation with a varying number of active expressions, each monitoring an index of the array, normalized by the interpretation median. Certain benchmark configurations are omitted due to scaling issues. Full results are given in table 6.

use different numbers of active expressions to monitor the value at specific instances of the array. We attach 10 no-op callbacks to each of those active expression.

**Discussion**   As shown in table 6, the compilation strategy implicates a high initial performance impact. According to the previous benchmark, this high impact results from the used source code transformation. However, with a growing number of active expressions the compilation strategy closes up to the interpretation strategy up to a point, where none is significantly faster. By increasing the number of active expressions even further, the relative slowdown of compilation compared to interpretation increases again, as seen in figure 10. The reason for this slowdown is that compilation uses a centralized update mechanism, in which all member accesses and associated active expressions are stored in a single data structure. In contrast, interpretation holds its associations separately in the property accessor of each member. Thus, the interpretation strategy scales better for higher numbers of active expressions.

## C   Implementing Reactive Concepts

In the following, we present implementations for the three remaining reactive concepts introduced in section 2: signals, constraints, and implicit layer activation (ILA).[12] For each concept, we provide both, an active expression-based implementation and a reference implementation that does not use active expressions.

---

[12] We omit details not relevant in the context of this work.

■ **Table 6** Benchmark timings and relative slowdowns running quicksort with a varying number of monitored indices.

| Number of Expressions | timings [ms] | | slowdown |
| --- | --- | --- | --- |
| | Compilation | Interpretation | Compilation vs Interpretation |
| 0 | 6.39 | 0.06 | 98.35 [97.85–98.73] |
| 1 | 6.10 | 1.46 | 4.19 [3.40–4.36] |
| 10 | 5.87 | 3.30 | 1.78 [1.74–1.83] |
| 20 | 6.72 | 5.20 | 1.29 [1.27–1.31] |
| 30 | 7.46 | 7.14 | 1.05 [1.03–1.08] |
| 40 | 12.33 | 8.74 | 1.41 [1.14–1.73] |
| 50 | 9.50 | 10.60 | 0.90 [0.88–0.92] |
| 100 | 14.14 | 18.86 | 0.75 [0.74–0.76] |
| 150 | 19.97 | 27.65 | 0.72 [0.71–1.83] |
| 200 | 28.79 | 35.15 | 0.82 [0.76–0.83] |
| 250 | 66.90 | 44.60 | 1.50 [0.76–1.80] |
| 300 | 82.77 | 48.98 | 1.69 [0.99–1.70] |

**C.1 Signals**

Signals are time-varying values that introduce functional dependencies in a program. In particular, signals update automatically whenever their dependencies change. For proper language integration, we apply a small syntax extension for signals to the underlying base language [31]. Signals are defined by a new type of VariableDeclaration called signal:

```
signal s = expr;
```

As this new syntax element is not defined in the base language, we first perform a simple source-to-source transformation of the signal definition into parsable JavaScript code. Then, each of the two implementations[13,14] applies an AST transformation to preserve the intended semantics as described in the following.

**Non-Active Expression-based Variant**   Instead of lifting every function to deal with both, primitive values and signals, we represent signals using ordinary JavaScript variables. However, we maintain a separate data structure to keep track of signals and their dependencies.

---

[13] https://github.com/active-expressions/programming-signals-plain accessed on February 28th 2017, at commit b3c5728
[14] https://github.com/active-expressions/programming-signals-aexpr accessed on February 28th 2017, at commit 9d3ccf7

**Active Expressions**

To identify the signals in the system, we traverse the AST for VariableDeclaration nodes of the respective type. Next, we transform the identified signal declarations into calls to the function defineSignal:

| Before transformation | After transformation |
|---|---|
| ```
signal s = expr;
``` | ```
let _scope2 = {
  name: '_scope2'
};
let s = defineSignal(_scope2, 's', () => expr, () => s =
    expr);
``` |

We distinguish local variables such as the signal s by their name and scope. To make the scope computationally accessible, we inject explicit scope objects as exemplified in line 1 to 3. The defineSignal function uses the name and scope information to create a signal meta object for the variable s:

```
defineSignal = function (scope, name, init, solve) {
  let signal = new Signal(scope, name, init, solve);
  signals.push(signal);
  return signal.initialize();
}
```

After creating the signal meta object, we call its initialize method to store its dependencies in line 4. To do so, the method set a flag indicating dependency detection and then simply executes the given initialization expression. To identify the dependencies, we transform all accesses to local and global variables, and object properties as well as method calls in the program into immediately invoked function expressions. The following transformation exemplifies this process for the local variable a:

| Before transformation | After transformation |
|---|---|
| `a` | ```
((result, scope, name) => {
  getLocal(scope, name);
  return result;
})(a, _scope2, 'a')
``` |

In addition to returning the value of the variable, we add the given name-scope-combination as a dependency to the currently analyzed signal in the getLocal function. As a result, we identified all dependencies of the current signal.

Next, we have to detect state changes in the running program to react to them appropriately. To detect those changes, we replace all assignments to local and global variables, and object properties with immediately invoked function expressions. In the following, we exemplify this transformation for an assignment to the local variable a:

| Before transformation | After transformation |
|---|---|
| `a = 42;` | ```
((result, scope, name) => {
  setLocal(scope, name);
  return result;
})(a = 42, _scope2, 'a');
``` |

After applying the assignment, we notify the reactive system about the change by calling the setLocal function. This function is responsible for updating the signal





network accordingly. Once we detect a change of an ordinary variable or an object property, the setLocal function checks whether any signal is affected by this change. To do so, we scan the list of signals for a signal that lists the changed name-scope-combination as a dependency. If such a signal exists, we update all signals that are affected by the change in topological order. Some imperative concepts, including if-statements and nested structures, may invalidate the current dependencies. Thus, we re-compute the dependencies of all updated signals. With the change propagated through the dependency graph, the signal network is consistent again and normal execution resumes.

**Active Expression-based Implementation**   Similar to the reference implementation, we use ordinary JavaScript variables to represent signals. In this implementation, we use Active expressions to keep track of all variables defining the value of a signal.

Again, we identify all signals in the system by traversing the AST for VariableDeclaration nodes of the respective type. Then, we replace any occurrence of a signal declaration with a SequenceExpression:

| Before transformation | After transformation |
|---|---|
| **signal** s = expr; | **let** s = (**aexpr**(() => expr).onChange(resolveSignals), ↪ signals.push(() => s = expr), expr); |

The generated SequenceExpression handles three tasks. First, we set up an active expression to monitor the signal expression for changes. When the evaluation result of the expression changes, we have to resolve the signal network. Second, for later usage, we store a resolving function. This function updates the signal according to the current values of its dependencies by re-assigning the expression to the local reference. Third, we initially assign the current value of the init expression to the signal variable.

Once we detect a change in a primitive value using active expressions, we update all signals in topological order. However, we have to consider other active expressions to prevent glitches. Glitches describe the situation of accessing a signal when the signal network is in a temporary inconsistent state because it currently updates [3]. In terms of active expressions, such glitches might happen if other active expressions depend on signals or their dependencies:

```
 1  let a = 0
 2  signal b = a;
 3  signal c = a + 1;
 4
 5  function checkConsistency() {
 6    if(c !== b + 1) throw new Error('subject to the glitch problem!');
 7  }
 8
 9  aexpr(() => b).onChange(checkConsistency);
10  aexpr(() => c).onChange(checkConsistency);
11
12  a++;
```



**Active Expressions**

To prevent these glitches, we defer the invocation of all callbacks of normal active expressions until the signal graph is updated completely. To do so, we wrap all other active expressions to be able to wait for the signal graph to fully update.

**Discussion**   In the reference implementation, we have to take care of dependency detection ourselves. Thus, we have to deal with many implementation details that come with the integration of signals with imperative environment. For instance, dependencies might become outdated, if our expression involves if-statements, nested data structures, or function calls. In these cases, dependencies have to be re-evaluated. Furthermore, we have to detect changes to appropriately react to them. However, this task involves various different types of AST nodes to be considered: assignments to local variables, assignments to `MemberExpressions`, and `UpdateExpressions` have to be handled differently. For example, to distinguish assignments to local variables of the same name, we had to inject explicit scope objects as an additional characteristic.

In contrast, active expressions provide expressions as a simple abstraction over imperative and OO concepts. Thus, active expressions handle state changes in this highly complex language for us. Without having to care about those tedious implementation details, programmers may focus on the reactive part of their implementations.

### C.2  Constraints

A common real-world use case for constraint programming is layouting. Layout constraints are described in a declarative manner and solved using specialized solvers such as the linear constraint solver Cassowary [2]. Constraint solvers are usually provided as a library. Integrating constraint solving libraries into an OO program typically leads to boilerplate code for setting up constraints and triggering solving behavior. To mitigate this verbosity, it is desirable to provide some syntactic sugar for constraint specification to integrate well with the existing imperative language and ensure familiarity for OO developers via familiar syntax. To exemplify an implementation of constraints in an OO scenario, we provide a reactive version of Cassowary, first without active expressions,[15] then with active expressions.[16] For our constraint syntax, we heavily draw on the notation proposed for BabelsbergJS [9]. Constraints are defined by a `LabeledStatement` with the label always:

```
always: a + b == c;
```

**Reactive Constraint Solving**   Cassowary cannot reason about ordinary JavaScript variables, but only about specialized `ConstraintVariables`. Thus, we apply a source-to-source transformation to transform the syntax described above into executable source code that fulfills the desired semantics. This transformation has three main tasks:

---

[15] https://github.com/active-expressions/programming-constraints-plain accessed on February 28th 2017, at commit 0af14a5
[16] https://github.com/active-expressions/babel-plugin-always-constraint accessed on February 28th 2017, at commit 4685a25





- Replace variables used in constraints with ConstraintVariables
- Handle accesses to those variables to make surrounding code unaware of dealing with special kinds of variables
- Lift the constraint expression to construct an actual constraint

To fulfill the first task, we have to identify all variables to be replaced. To do so, we first traverse the AST for constraint expressions, namely LabeledStatements with the label always. Next, we traverse these expressions for all Identifiers referenced in the expressions. For those variables, we compute the corresponding binding. Using the bindings, we identify the original VariableDeclarations for these variables. Instead of initializing the variable with an ordinary JavaScript value, we transform the VariableDeclaration to use ConstraintVariables:

| Before transformation | After transformation |
| --- | --- |
| `var a = 3;` | `var a = _newConstraintVar("a", 3);` |

When initializing a variable, the newConstraintVar function asks a shared solver instance to create an appropriate constraint variable.

The second task requires us to transform all references to constraint variables. Using the already computed bindings, we can identify all read accesses to constraint variables. We transform the plain access to these references to calls of the value method of the constraint variable:

| Before transformation | After transformation |
| --- | --- |
| `console.log(a);` | `console.log(a.value());` |

For write accesses, we identify all Identifier that are left hand of an AssignmentExpression and belong to a binding of a constraint variable. We rewrite the corresponding AssignmentExpression into a call of the setConstraintVar function:

| Before transformation | After transformation |
| --- | --- |
| `a += 42;` | `_setConstraintVar(a, "+=", 42);` |

The setConstraintVar function first determined the new target value of the constraint expression. Then, it creates a temporary soft equality constraint between the variable and the target value. Next, we ask the solver instance to solve the complete constraint system before removing the temporary constraint again. Solving the constraint on assignments to a variable makes the library reacting to change.

For the third task, we lift the expression of an always statement to appropriate function calls of the constraint variables. For example, we transform the + and the == operator into calls to the plus and the cnEquals member function, respectively. The addConstraint function adds the created constraint to the constraint system of the solver instance and solves the system:

| Before transformation | After transformation |
| --- | --- |
| `always: a + b == c;` | `_addConstraint(a.plus(b).cnEquals(c));` |

**Active Expression-based Implementation**   In contrast to the previous variant, we do not transform every variable referenced in a constraint expression. Instead, we only



### Active Expressions

◼ **Listing 1** The constraint **always**: a + b == c; in its rewritten form

```
1  {
2    let solver = Cassowary.ClSimplexSolver.getInstance();
3  
4    let _constraintVar_a = solver.getConstraintVariableFor(_scope, "a", () => {
5      let _constraintVar = new Cassowary.ClVariable("a", a);
6      aexpr(() => a).onChange(val => _constraintVar.set_value(val));
7      aexpr(() => _constraintVar.value()).onChange(val => a = val);
8      return _constraintVar;
9    });
10 
11   /*
12    * Analogous initializers for
13    * _constraintVar_b
14    * _constraintVar_c
15    */
16 
17   let linearEquation = _constraintVar_a.plus(_constraintVar_b).cnEquals(_constraintVar_c);
18   solver.addConstraint(linearEquation);
19 
20   trigger(aexpr(() => _constraintVar_a.value() + _constraintVar_b.value() == _constraintVar_c.
       ↪ value()))
21     .onBecomeFalse(() => solver.solveConstraints());
22 }
```

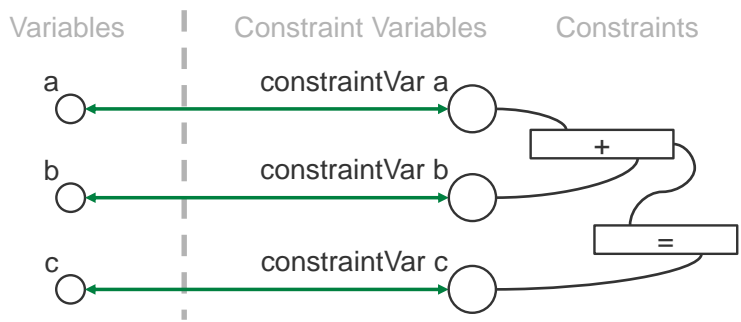

◼ **Figure 11** The constraint **always**: a + b == c; creates a constraint variable for each variable referenced by the constraint. Active expressions keep them in sync.

transform the constraint expression itself. For instance, the constraint **always**: a + b == c; would be transformed into the source code in listing 1. The transformed code snippet achieves the desired semantics by the following steps:

1. We get a **shared solver instance** to work with, as shown in line 2.
2. For each referenced variable, we ask the solver for a corresponding constraint variable in line 4. If the constraint variable does not already exists, we have to create a new **constraint variable**, as exemplified in line 5 to 8. To distinguish





variables of the same name in different scopes, we inject an explicit scope object into the scope of that variable and reference it here. Constraint variables act as a stand-in for the ordinary variables during constraint construction and solving. Thus, they should mirror the value of the ordinary variable. Therefore, we to introduce a *two-way data binding* between the two variables: when the value of one variable changes, we automatically adjust the other using the two active expressions in line 6 and 7. As a result, imperative JavaScript computations operate on ordinary variables and the constraint solver deals the constraint variables as dedicated representations, while keeping them in sync, as shown in figure 11.

3. As with the previous variant, we have to **lift the constraint expression** into appropriate library calls of the generated constraint variables, as shown in line 17. Next, we add the lifted constraint to the constraint system to be satisfied by the solver.

4. Because the Cassowary library does not detect invalidated constraints automatically, we use an active expression to **detect unsatisfied constraints**, as seen in line 20. For convenience, we again use *triggers* to reason about the expression result. Any callback given to the `onBecomeFalse` method is executed, whenever the expression result becomes false (from a non-false value) or if the expression result is initially false. As shown in line 21, whenever we detect an unsatisfied constraint, we use the solver to satisfy the constraint system. After solving the constraint system, the solver applies changes to the constraint variables. These changes propagate automatically to the ordinary JavaScript variables thanks to the two-way data bindings.

**Discussion**  This implementation exemplifies how active expressions may interface with complex reactive behavior. In contrast to the reference implementation, the active expression-based implementation does not rely on implicit lifting of variable accesses. Instead, the active expression variant introduces a clean separation between ordinary variables and their constraint equivalents. Using two-way data bindings, we can easily propagate changes from one paradigm to the other, thus, keeping both world views in sync. By taking care of change detection, active expressions help the concept programmer to focus on interfacing with the constraint library.

**C.3  Implicit Layer Activation**

As our final example, we provide an implementation of implicit layer activation (ILA) [21], an activation means for *layers* in context-oriented programming (COP). As described in section 2, layers are units of modularity that encapsulate adaptations to the program behavior. To apply these adaptations to the program, one activates a layer using one of multiple activation means. For ILA, a layer is active if and only if a given condition holds. Currently, ContextJS [20] does not support ILA as an activation means.[17] Thus, we show how to extend ContextJS with ILA, first using a non-reactive implementation, then with active expressions.

---

[17] https://github.com/LivelyKernel/ContextJS accessed on February 16th, at commit 938e117; npm package: contextjs in version 2.0.0



**Active Expressions**

**Imperative Implementation**  ContextJS supports multiple activation means, including global activation and dynamic activation for the extent of a function call. When calling a layered method, the currentLayers function is responsible for computing an appropriate layer composition, either by using a cached result or, if necessary, by determining a new one using global and dynamic layers:

```
export function currentLayers() {
  // parts omitted for readability
  if (!current.composition) {
    current.composition = composeLayers(LayerStack);
  }
  return current.composition;
}
```

For our extension,[18] we add a separate list of layers, called implicitLayers, to represent layers potentially activated through ILA. To implicitly activate a layer, we add the method activeWhile to the class Layer:

```
activeWhile(condition) {
  if (!implicitLayers.includes(this)) {
    implicitLayers.push(this);
  }
  this.implicitlyActivated = condition;

  return this;
}
```

This method has two responsibilities. First, it adds the layer to implicitLayers, the list of implicitly activated layers, in line 3 if necessary. Second, it stores the given condition under which the layer should be active, as seen in line 5. Using the list of implicitly activated layers, we can get all layers that are actually activated by filtering this list for layers with their conditions evaluating to true, as done by the getActiveImplicitLayers function:

```
function getActiveImplicitLayers() {
  return implicitLayers.filter(layer => layer.implicitlyActivated());
}
```

Using the getActiveImplicitLayers function, we can now adjust the computation of the current layer composition showed earlier:

```
export function currentLayers() {
  // part omitted for readability
  var current = LayerStack[LayerStack.length — 1];
  if (!current.composition) {
    current.composition = composeLayers(LayerStack);
  }
  return current.composition.concat(getActiveImplicitLayers());
}
```

To include implicitly activated layers, we append all layers activated through ILA to

---

[18] https://github.com/active-expressions/programming-contextjs-plain accessed on February 16th 2016, at commit 22deb54





the already computed layer composition. As a result, the returned layer composition contains dynamically activated, globally activated, and implicitly activated layers. Note, that we cannot cache implicitly activated layer, because we have no means to invalidate the cache on changes to the condition.

**Active Expression-based Implementation**   In contrast to the imperative implementation, we do not introduce a separate data structure for implicitly activated layers. Instead, we treat implicitly activated layers when their condition evaluates to true as being globally active. As a result, we can reuse the existing layer composition algorithm once a layer becomes active. To do so, the activeWhile method has to setup dependencies to detect changes to the given condition and update the layer accordingly. For this implementation,[19] we assume the parameter condition to be provided as an active expression. Using this active expression, we can implement the appropriate reactive behavior:

```
activeWhile(condition) {
  trigger(condition)
    .onBecomeTrue(() => this.beGlobal())
    .onBecomeFalse(() => this.beNotGlobal());

  return this;
}
```

As with reactive object queries and constraints, we use triggers for convenience in line 2. Whenever the expression result becomes true or false, the layer is activated or deactivated in line 3 or 4, respectively. Additionally, the trigger behavior automatically adjusts the initial state of the layer depending on the current result of the given expression.

**Discussion**   Usually, implicit layer activation (ILA) does not require a reactive implementation. Instead, the current layer composition stack is determined imperatively when calling a layered method. At this very point in time, the COP framework checks the conditions of all implicitly activated layers. However, this non-reactive approach is only possible, because most COP implementations are limited to adapting object and class methods [25]. These concepts are *passive* entities that only affect the program behavior when called explicitly. Therefore, the limitation to passive entities enables the COP framework to check the condition at a well-defined point in the program. In contrast, *active* entities, such as constraints, may initiate behavior by themselves without being called explicitly. Extending the concept of COP beyond method decoration requires to activate scoped entities at specific times. The presented reactive implementation is not significantly simpler than the imperative one, however, it enables a COP framework to deal with active elements. Thus, a reactive implementation paves the way to apply the concept of COP to other types of abstraction beyond partial methods. For example, a layer could be used to limit the scope of a constraint: when

---

[19] https://github.com/active-expressions/programming-contextjs-aexpr accessed on February 16th 2016, at commit 07437e8



**Active Expressions**

a condition becomes true, the corresponding layer becomes active and the constraint immediately takes effect, instead of waiting for an additional, explicit trigger [18].

## D  Source Code for Trigger

Triggers are a small utility library on top of active expressions to better reason about change in expression results. In particular, trigger ease working with expressions that have boolean evaluation results.

```
1  class Trigger {
2    constructor(aexpr) {
3      this.aexpr = aexpr;
4    }
5
6    onBecomeTrue(callback) {
7      this.aexpr.onChange(bool => {
8        if(bool) callback();
9      });
10     if(this.aexpr.now()) callback();
11
12     return this;
13   }
14
15   onBecomeFalse(callback) {
16     this.aexpr.onChange(bool => {
17       if(!bool) callback();
18     });
19     if(!this.aexpr.now()) callback();
20
21     return this;
22   }
23 }
24
25 export default function trigger(aexpr) {
26   return new Trigger(aexpr);
27 }
```

For example, the onBecomeTrue method from line 6 to 13 invokes a given callback, whenever the result of the given expression becomes true from a non-true value. Additionally, triggers check whether the expression evaluates to true initially, as shown in line 10. The onBecomeFalse method works analogously.

## E  Code Complexity Measurement Methods

To quantitatively compare the respective implementations in section 5 and appendix C, we analyze them regarding code complexity. As a metric for code complexity, we measure the number of AST nodes in the implementations as described in the following:

12-46



**Code Analysis**   To generate an AST from the source code to analyze, we use the babylon[20] parser as npm package at version 6.16.1 with the following parsing options:

```
sourceType: "module",
plugins: [
    "estree",
    "jsx",
    "flow",
    "doExpressions",
    "objectRestSpread",
    "decorators",
    "classProperties",
    "exportExtensions",
    "asyncGenerators",
    "functionBind",
    "functionSent",
    "dynamicImport"
]
```

Then, we determine the number of AST nodes by traversing the AST and counting all nodes except comment types. Finally, we aggregate the results of all files.

**Analyzed Implementations**   For our analysis, we analyzed all source code files transitively referenced by the entry point of the project, excluding libraries, tests and files concerning workflow. We analyzed the following projects:

- **Signals, plain:** https://github.com/active-expressions/programming-signals-plain.git at commit b3c5728
- **Constraints, plain:** https://github.com/active-expressions/programming-constraints-plain.git at commit 0af14a5
- **Reactive Object Queries, plain:** https://github.com/active-expressions/programming-roq-plain.git at commit 0cad577
- **Implicit Layer Activation, plain:** https://github.com/active-expressions/programming-contextjs-plain.git at commit 1360e1a
- **Signals, Active Expressions:** https://github.com/active-expressions/programming-signals-aexpr.git at commit 9d3ccf7
- **Constraints, Active Expressions:** https://github.com/active-expressions/babel-plugin-always-constraint.git at commit 4685a25
- **Reactive Object Queries, Active Expressions:** https://github.com/active-expressions/reactive-object-queries.git at commit f45fe5e
- **Implicit Layer Activation, Active Expressions:** https://github.com/active-expressions/programming-contextjs-aexpr.git at commit 07437e8

For both signal implementations, we preprocessed the entry file of the project by extracting the setup template into a separate file because this string semantically represents source code, in particular the change propagation of the concept. We added

---

[20] https://github.com/babel/babylon accessed on February 25th 2017



**Active Expressions**

implicit layer activation (ILA) as a new activation means to ContextJS, an existing COP implementation. Thus, we measure only the additional complexity added to ContextJS by our two implementations of ILA. To do so, we determine the same complexity metric for ContextJS[21] and calculate the difference to both implementations with ILA.

---

[21] https://github.com/LivelyKernel/ContextJS.git at commit 938e117





**About the authors**

**Stefan Ramson** is a doctoral researcher at the Software Architecture Group. His research interests include programming language design and natural programming. Contact Stefan at stefan.ramson@hpi.uni-potsdam.de.

**Robert Hirschfeld** is a Professor of Computer Science at the Hasso Plattner Institute at the University of Potsdam, Germany. His Software Architecture Group is concerned with fundamental elements and structures of software. Contact Robert at hirschfeld@hpi.uni-potsdam.de.